\documentclass{article}

\usepackage{arxiv}

\usepackage{latexsym}
\usepackage{graphicx}
\usepackage{multicol,multirow}
\usepackage{amsmath,amssymb,amsfonts}
\usepackage{mathrsfs}
\usepackage{amsthm}
\usepackage{algorithmic}
\usepackage{algorithm}
\usepackage{apacite}
\usepackage{rotating}
\usepackage{appendix}
\usepackage[authoryear]{natbib}
\usepackage{ifpdf}
\usepackage[T1]{fontenc}
\usepackage{times}
\usepackage{sourcesanspro}
\usepackage{newtxmath}
\usepackage{textcomp}%
\usepackage{xcolor}%
\usepackage{hyperref}
\usepackage{caption}
\usepackage{subcaption}

\usepackage[utf8]{inputenc} % allow utf-8 input
\usepackage[T1]{fontenc}    % use 8-bit T1 fonts
\usepackage{url}            % simple URL typesetting
\usepackage{booktabs}       % professional-quality tables
\usepackage{nicefrac}       % compact symbols for 1/2, etc.
\usepackage{microtype}      % microtypography
\usepackage{lipsum}		% Can be removed after putting your text content

\usepackage{lineno}
\usepackage{soul}
\usepackage{dcolumn}
\usepackage{colortbl} %use in the preamble
\newcolumntype{L}{D{.}{.}{4,1}}
\newcommand{\mb}[1]{\boldsymbol{#1}}
\newcommand{\bm}[1]{\boldsymbol{#1}}
\DeclareMathOperator*{\argmax}{arg\,max}
\DeclareMathOperator*{\argmin}{arg\,min}
\colorlet{shadecolor2}{gray!20}
\definecolor{shadecolor}{RGB}{144,238,144}

\defcitealias{AIS-IMO:2003}{IMO, 2003}

\title{Autocalibration of the E3SM Version 2 Atmosphere Model Using a PCA-Based Surrogate for Spatial Fields}

\author{Drew Yarger\\
	Sandia National Laboratories\\
	Albuquerque, NM \\  \\
	Purdue University\\
	West Lafayette,Indiana\\
	\And
	Benjamin Wagman\\
	\thanks{corresponding author,bmwagma@sandia.gov} \\
	Sandia National Laboratories\\
	Albuquerque, NM \\
	\And
	\href{https://orcid.org/0000-0002-1731-3834}{ \includegraphics[scale=0.06]{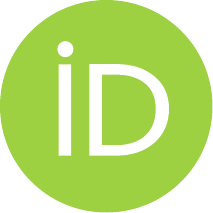}\hspace{1mm}Lyndsay Shand} \\
	Sandia National Laboratories\\
	Albuquerque, NM \\	\\
	University of Illinois, Urbana-Champaign\\
	Urbana, Illinois\\
	\And
	Kenny Chowdhary\\
	NVIDIA\\
	Santa Clara, CA \\
}

\renewcommand{\shorttitle}{E3SMv2 Autotuning}

\begin{document}
\maketitle

\begin{abstract}
Global Climate Model (GCM) tuning (calibration) is a tedious and time-consuming process, with high-dimensional input and output fields. Experts typically tune by iteratively running climate simulations with hand-picked values of tuning parameters. Many, in both the statistical and climate literature, have proposed alternative calibration methods, but most are impractical or difficult to implement. We present a practical, robust, and rigorous calibration approach on the atmosphere-only model of the Department of Energy’s Energy Exascale Earth System Model (E3SM) version 2. Our approach can be summarized into two main parts: (1) the training of a surrogate that predicts E3SM output in a fraction of the time compared to running E3SM, and (2) gradient-based parameter optimization. To train the surrogate, we generate a set of designed ensemble runs that span our input parameter space and use polynomial chaos expansions on a reduced output space to fit the E3SM output. We use this surrogate in an optimization scheme to identify values of the input parameters for which our model best matches gridded spatial fields of climate observations. To validate our choice of parameters, we run E3SMv2 with the optimal parameter values and compare prediction results to expertly-tuned simulations across 45 different output fields. This flexible, robust, and automated approach is straightforward to implement, and we demonstrate that the resulting model output matches present day climate observations as well or better than the corresponding output from expert tuned parameter values, while considering high-dimensional output and operating in a fraction of the time.
\end{abstract}

% keywords can be removed
%\keywords{cloud-aerosol interactions \and GOES-R \and ship tracks}

\section{Introduction}

Model tuning, also referred to as model calibration, is the science (and art) of matching the model predictions to observed data by adjusting model parameters after the model configuration has been fixed \cite{hourdin_2017}.
Improvements in model tuning can contribute to improvements in model predictions of the past, present, and future, and a better understanding of uncertainty. 

Earth Systems Models (ESMs) are some of the most computationally-expensive and complex computer models requiring tuning. 
When tuning ESMs, the goal is to obtain model output that is consistent with our observations of the Earth; in many cases, these observations consist of different time-averaged spatial fields, which may have hundreds or thousands of spatial grid points.
In this setting, the input parameters govern aspects of physical processes and typically are on the order of 10's to 100's for an ESM. For tuning, the set of adjusted model parameters is down-selected to 10's of parameters or less to make the problem tractable.
Traditionally, ``hand-tuning" for Earth System Models is a slow and tedious process in which experts select a set of parameters, manually adjusts them, and evaluates their effects on the simulation. The difficulty, in part, is due to the aforementioned complexity of the models. 
Furthermore, ``hand-tuning" is not formalized by a set of rules, so a different expert will likely derive a different solution.

Alternative approaches to model tuning by hand have been and continue to be proposed, but many are still challenged by scalability. %the high dimensionality of the problem space.
For example, \cite{jackson2004efficient} and \cite{ jackson2008error} proposed the Multiple Very Fast Simulated Annealing Algorithm to efficiently identify the regions of the parameter space that minimize differences between predictions and observations within a Bayesian framework. This, however, still relies on evaluating the computer model at various parameter values sequentially, rather than simultaneously, which leads to a slower tuning process. An increasingly popular approach, due to the surge in machine learning tools, is to build an ``emulator'' or ``surrogate'' which maps the input parameter space to the output space using a perturbed parameter ensemble. Apart from the generation of the training data, which is the most time-consuming part of the process, the emulator or surrogate can typically run orders of magnitude faster. Whereas a single E3SM simulation run takes days to run, a surrogate model can typically produce output in a fraction of a second. This is the approach we will take here.

In the statistical literature, a classical calibration solution \cite{kennedy2001bayesian} uses a Gaussian Process (GP) emulator within a Bayesian framework to account for known sources of uncertainty, e.g., discrepancy between the surrogate and the climate model, or between the climate model and observational data.
%in which they propose Gaussian process (GP) emulators within a Bayesian framework to account for known sources of uncertainty (e.g.\ both climate and surrogate model discrepancies).
Their method launched many proposed GP-based calibration approaches, yet it was employed in a simplified setting and requires computationally-expensive iterative retraining of the surrogate model. % \note[DY]{Previous version commented out, tried to edit with a different flavor} \note[BMW]{Better, but please clarify: is it too expensive to apply to ESM's? }\note[LS]{it requires continual retraining making the model fitting process expensive.}
\cite{higdon2013computer} demonstrate a multi-step ensemble Kalman filtering approach for high-dimensional model output, which also leverages GP emulators and accounts for model discrepancy from the Community Atmosphere Model (CAM). 
\cite{dunbar2021calibration}, \cite{berdahl2021statistical}, and \cite{beusch2021emission} have all utilized a Gaussian process emulator approach for the calibration of an idealized GCM and for the Community Ice Sheet Model (CISM), respectively. 
The ``Calibrate, Emulate, Sample'' approach of \cite{cleary2021calibrate} leverages GPs as emulators within an approximate Bayesian learning framework to ease the computational burden of traditional Bayesian frameworks. Recently, proposed autocalibration approaches are also leaning into approaches with more of a machine learning flavor. 
\cite{fletcher2022toward} proposes a convolutional neural network as the emulator for multiple spatial fields, but this has not yet been implemented fully within a calibration framework. 
For a thorough literature review of climate model emulators that could be used for calibration, see Chowdhary et al. (2021). 

%\note[LS]{this paragraph could more concisely state the novelty of our method to further cut down the intro. i  made some suggestions below. i still feel we could do a better job clarifying the novelty of our approach}
Building on this body of work, we propose 
a flexible and scalable framework for the automated calibration of the atmospheric model of E3SM, which accounts for uncertainty in the parameter space and spatial variability of high-dimensional time-averaged spatial fields. Similar to the approaches above, we rely on a surrogate model for calibration. Instead of the popular GP surrogates, we propose to use polynomial chaos expansions (PCE), as in \cite{ricciuto2018impact}, on a reduced space.
%\note[LS]{we could move the discussion of why we built on a reduced space (sentences in blue) to the methods section} \note[DY]{I lik that}
Our approach fills a need for more automated training tools and meta-learning approaches for surrogate construction that can adapt to data-rich or data-sparse environments. The general calibration framework proposed allows the user to select their choice of surrogate model (i.e.,\ alternatives to PCE) and loss function and is intended to be easy to implement and understandable by any user. 
In the case of a very sparse data setting, it is possible and recommended to combine expert decision making with this approach for the greatest impact. 

Our proposed framework is applied to the atmosphere model of E3SMv2, resulting in an a optimized set of 5 parameters that have substantial influence on the 45 output fields we studied. 
We validate these parameters using an E3SMv2 simulation, and find that the optimized set results in a simulated climate that reduces the root-mean-squared-error (RMSE) relative to our observational targets for a majority of our targeted fields, as compared to hand-tuned E3SMv2. Many biases of the hand-tuned E3SMv2 persist in the autocalibrated E3SMv2; these could likely be reduced by tuning a larger set of E3SM parameters or by improving the structure of the E3SM model.
The following is a brief outline of the paper. Section \ref{sec:data} describes how we generated the training data for our surrogate, section \ref{sec:method} steps through the surrogate model fitting and inverse optimization over a loss function, and section \ref{sec:results} demonstrates our approach in the context of E3SMv2. Lastly, in section \ref{sec:disc} we discuss special considerations and future directions of our approach.

\section{Perturbed Parameter Ensemble}\label{sec:data}

%\subsection{Climate model, parameters, and target fields}\label{sec:e3sm_info}

To train our surrogate,  we generate an ensemble of 250 10-year simulations using the Energy Exascale Earth System Model version 2 (E3SMv2) \citep{e3smv2overview}. The simulations are configured with active atmosphere, land, and river components, while ocean surface temperature and sea ice extent are prescribed to a monthly-varying present-day observational climatology for years 2005 to 2014 \cite{sstdata}. This configuration is commonly referred to as ``atmosphere only" even though the land and river models are active. The atmosphere model has approximately 110 km grid spacing and 72 vertical levels with a model top of 10 Pa (approximately 60 km). The prescribed ocean configuration is consistent with the goal of finding atmospheric parameters that bring the climate closest to the present-day atmospheric targets while eliminating the influence of ocean variability and minimizing the time necessary for the simulated climate to adjust to parameter perturbations.

Five atmospheric input parameters (see table \ref{tab:parameters}) were adjusted using Latin Hypercube Sampling \cite{McKay1979}. The selected parameters come from the shallow convection and stratiform cloud parameterization \cite{Golaz2002clubb, larson_clubb}, the deep convection parameterization \cite{zm95, xie2019}, and the cloud microphysics parameterization \cite{MG2}. Parameterizations in ESM's represent one part of a simplification of a physical process that occurs in the climate system but is not resolved in the model. These simplifications often involve empirical fits or bulk representations of populations, so uncertainties arise naturally from using a single number to approximate these complexities. The five selected parameters and their sampling bounds were chosen in consultation with E3SM developers who previously ``hand-tuned" E3SMv2. Each of the five parameter's influence on model output is demonstrated on an E3SMv1 perturbed parameter ensemble \cite{Qian2018}. By restricting the ensemble to five parameters we achieve relatively dense sampling, but we knowingly exclude other influential parameters. Including more parameters could result in a better calibration. 

\begin{table}
    \centering
    \caption{Atmospheric parameters sampled. ``pdf" is probability density function, and CAPE is convective available potential energy.} \label{tab:parameters}
    \begin{tabular}{lllll}
        \hline
       Short Name & Description & Low & High  \\ \hline  
       \texttt{ice\_sed\_ai} & Fall speed parameter for cloud ice & 350 & 1400  \\
       \texttt{clubb\_c1} & Const. for dissipation of variance of vertical wind ($\overline{w'^2}$) & 1.0 & 5.0   \\
       \texttt{clubb\_gamma\_coef} & Const. of width of pdf in $w$ coord. & 0.10 & 0.50  \\
       \texttt{zmconv\_tau} & Time scale for consumption rate of deep CAPE (s) & 1800 & 14400  \\
       \texttt{zmconv\_dmpdz} & Parcel fractional mass entrainment rate & -2.0e-3 & -0.1e-3 \\
    \end{tabular}
\end{table}

The target fields of interest are time-averaged (over the 10-year period) for every season and spatial location of eleven climate variables, creating a total of 44 target spatial fields. These, along with a scalar target value, \texttt{RESTOM}, are described in table \ref{tab:fields}.
We primarily use simulation and observational data that has been coarsened from its original resolution to a 24x48 grid (7.5 degree resolution, 1,152 grid points) resolution for latitude and longitude fields and 24x37 (888 grid points) resolution for latitude and pressure fields. 
We found that working with this resolution worked well in our setting and minimized the influence of more extreme observations. 
The method can easily be extended for finer scale resolutions, and our implementation allows a user to select 180x360 (1 degree resolution, 64,800 grid points) and 180x37 (6,660 grid points) spatial fields. 

    \begin{table}
    \centering
    \caption{Target fields.} \label{tab:fields}
    \begin{tabular}{lllll}
        \hline
       Short Name & Description & Dimension & Reference  \\ \hline  
       LWCF & Longwave cloud forcing  & lat x lon & ceres\_ebaf\_toa  \\
       SWCF & Shortwave cloud forcing  & lat x lon & ceres\_ebaf\_toa     \\
       PRECT & Precipitation  & lat x lon & GPCP1DD   \\
       PSL & Sea level pressure  & lat x lon & ERA Interim   \\
       U200 &Zonal wind at 200 mbar pressure surface & lat x lon & ERA Interim  \\
       U850 & Zonal wind at 850 mbar pressure surface  & lat x lon & ERA Interim \\
       Z500 & Geopotential height at 500 mbar pressure surface  & lat x lon & ERA Interim\\
       TREFHT & Reference height temperature & lat x lon& ERA Interim  \\ 
       U & Zonal wind   & lat x plev& ERA Interim\\
       T & Temperature  & lat x plev& ERA Interim\\
       RELHUM & Relative humidity  & lat x plev & ERA Interim \\
       RESTOM & Net radiation flux at the model top & global & N/A  \\ 
    \end{tabular}
\end{table}

The E3SMv2 PPE was generated on the Chrysalis supercomputer at Argonne National Laboratory. Parameter sampling and E3SMv2 model setup was managed using the uncertainty quantification software Dakota \cite{dakota}. Model throughput is often a bottleneck for PPE's, but computational resources enabled relatively fast ensemble generation. Node counts were varied experimentally before settling on ten nodes per simulation, each node having 64 cores.  Simulations were submitted ten-at-a-time in 100-node bundles. These ten-simulation bundles achieved $\sim$90-100 simulated years per day (SYPD), and were sometimes run two-at-a-time for a combined $\sim$180-200 SYPD. The ensemble consists of 2,500 simulated years, so such an ensemble could be run, theoretically, on 200 dedicated nodes in about 14 days. We emphasize throughput because ensembles of future E3SM versions will be planned in parallel with model development to assist with hand tuning, and such ensembles must be produced quickly to be relevant. 

\section{Method}\label{sec:method}
%\note[LS]{i didn't like how vague it was to refer to ``optimal" parameters so many times so i just defined these to be theta-hat here and use that below.}

This section gives a description of the framework we propose for automated calibration, including the surrogate fitting, optimization steps, and implementation tools used. 
Let $\mb{\theta} = (\theta_1, \ldots, \theta_d)$ be a $d$-dimensional variable describing the atmospheric parameters and let $f(\mb{\theta}) \in \mathbb{R}^m$ denote the $m-$dimensional time-averaged spatial fields, where we unfold and stack the target outputs of the ESM into a single output vector. In our exemplar, $d=5$ as shown in table \ref{tab:parameters} and $m = \sum_{p} m_p$ where $m_p$ denotes the size of the spatial elements of the $p$-th field, for a total 47,404 elements containing data from four seasons after removing missing data (data with \texttt{NA} values). 
The goal of calibration is to find values for the vector of parameters $\mb{\theta}$ that minimize a loss function $L(\cdot, \cdot)$ between $f(\mb{\theta})$ and an $m$-dimensional vector of target observations $\mb{Y}_{obs}$, i.e.\ $L(f(\mb{\theta}), \mb{Y}_{obs}).$ 

Substituting $f(\mb{\theta})$ with a surrogate model, $\hat{f}(\mb{\theta},\mb{\phi}),$ we can write the solution, or ``optimal" parameter set, as
\begin{linenomath}
\begin{align}\label{eq:loss}
    \mb{\hat\theta} &= \argmin_{\mb{\theta}} L(\hat{f}(\mb{\theta}, \mb{\hat{\phi}}), \mb{Y}_{obs}).
\end{align}
\end{linenomath}
Here, while $\mb{\theta}$ refers to the input E3SM parameters that were perturbed, the parameters $\mb{\phi}$ refers to the surrogate-specific parameters (and $\mb{\hat{\phi}}$ to its estimate), e.g.,\ the polynomial coefficients in the case of a polynomial regression surrogate. % \note[BMW]{Confusing to me: are the surrogate parameters unique to the surrogate? I think so but it's not quite clear} \note[DY]{I'm not sure I understand. Will think about it} \note[DY]{I slightly reworded but don't really know if that's what you want}\note[LS]{i reworded to make more clear with an example}. 

The surrogate $\hat{f}(\mb{\theta}, \mb{\hat{\phi}})$ aims to mimic the behavior of $f(\mb{\theta})$ at a fraction of the computational cost.
The task of finding a set of estimated tuning parameters $\mb{\hat{\theta}}$ can then be broken out into two steps: 1) fitting the surrogate model $\hat{f}(\mb{\theta},\mb{\phi})$ to the training ensemble (i.e.\ learning $\mb{\hat{\phi}}$, an estimate of $\mb\phi$) and 2) solving equation \ref{eq:loss} with a specified loss function to learn $\mb{\hat\theta}$.

\begin{figure}
    \centering
    \includegraphics[width = .98\textwidth]{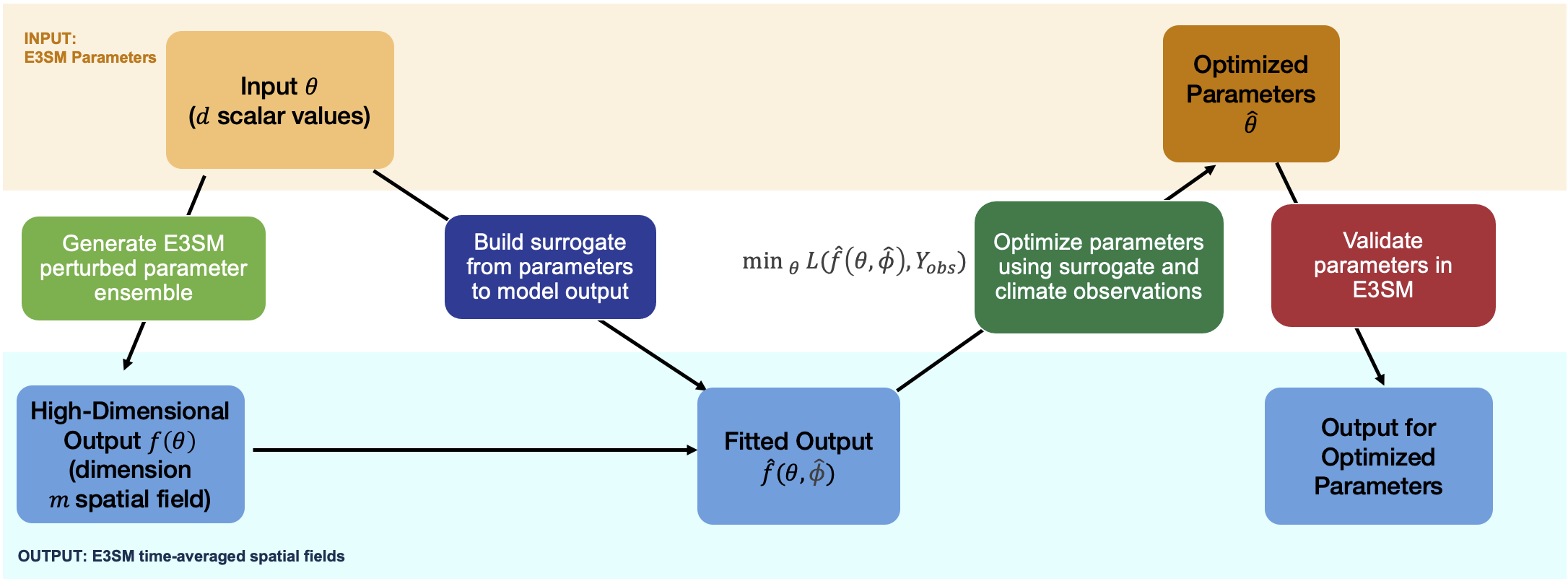}
    \caption{An overview of our automated calibration workflow broken into four main steps (depicted in center row): generation of PPE, estimation of surrogate model, optimization, and validation.}
    \label{fig:auto_framework}
\end{figure}

We detail the specifics of building the surrogate in section \ref{sec:pce_rom}, and we discuss the optimization step in section \ref{sec:optimization_methods}. 
A summary and visual outline of our automated calibration workflow is given in figure \ref{fig:auto_framework}. 

\subsection{Surrogate Model Construction}\label{sec:pce_rom}

As mentioned in section \ref{sec:data}, we use a perturbed parameter ensemble of size $n=250$ to train our surrogate model; we briefly detail the choice of surrogate and surrogate training process and provide links to open-source software implementation
here. 
Let $\mb{X} \in \mathbb{R}^{n \times d}$ be the perturbed parameter ensemble of input values theta, sampled via Latin Hypercube Sampling (LHS), and let $\mb{Y}\in \mathbb{R}^{n \times m}$ be the resulting E3SM model output of the target fields, such that the data matrix pair $(\mb{X}, \mb{Y})$ represents the full set of ensemble training data. 

Due to the high dimensionality of the target fields, i.e., $m \gg 1$, training $m$ independent surrogate models is impractical, especially when choices of hyperparameters (i.e.\ values describing the structure of the surrogate model) are to be chosen automatically. %\note[KC]{removed, but hyperparameter tuning is crucial to the success of our methodology.}\note[LS]{is hypertuning different than training the surrogate? i opt to remove what is in parentheses here if it gives no additional information.}. 
Further, treating target fields independently of others would ignore the inherent correlation across these variables. %and lead to scientifically unrealistic solutions. 
For a practical solution, we reduce the dimension of our output space via principal component analysis (PCA, a.k.a.\ empirical orthogonal function decomposition) and train a surrogate on each of the orthogonal components. This technique can also be referred to as data-driven Reduced Order Modeling (DDROM), the purpose of which is to reduce the output size of the quantities of interest \cite{Chowdhary2022}.
Specifically, we compute a set of principal components $\{\mb{\psi}_j \in \mathbb{R}^{m}\}_{j=1}^k$ and projection coefficients $\{\mb{\eta}_j \in \mathbb{R}^{n}\}_{j=1}^k$ from $\mb{Y}$ where $k \ll m$. 
This results in a rank-$k$ approximation of $\mb{Y}$: \begin{linenomath}\begin{align*}
    \mb{Y} \approx \sum_{j=1}^k \mb{\eta}_j \mb{\psi}_j^\top,
\end{align*}
\end{linenomath}where $\mb{\psi}_j^\top$ represents the transpose of the column vector $\mb{\psi}_j$.  
We then independently train separate surrogate models, $\hat{f}_j(\mb{\theta}, \mb{\phi}_j)$ to each of the $k$ projection coefficients using Legendre-based polynomial-based regression models, also known as polynomial chaos expansions (PCEs) \cite{Xiu2002, Xiu2005}. Now, 
instead of training $m$ regression models, we fit $k$ PCE-based machine-learning models
on $k$ training pairs, $(\mb{X}, \mb{\eta}_j)$ for $j=1,\dots,k$. 

The polynomial chaos expansion can be written as a linear combination of basis terms, 
\begin{linenomath}\begin{align*}
    \hat{f}_j(\mb{\theta}, \mb{\phi}_j) = \sum_{i=0}^P \phi_{ij}L_i(\bm{\theta}), 
\end{align*}
\end{linenomath}
where $L_i(\bm{\theta})$ represents the $d$-dimensional multivariate Legendre polynomials \cite[2.2.2]{Xiu2005}, which are products of its univariate counterpart \cite[4.2.1]{Hesthaven2007} along the $d$ dimensions.
The total number of expansion terms, $P+1$, is given by $(d+p)!/(d!p!)$,
where $p$ is the maximum polynomial order, 
which means that the number of terms can grow very quickly \cite{Xiu2002}. 
Nonetheless, we can efficiently compute these coefficients in a systematic, 
robust, and automated manner. We use $k$-fold cross validation and a brute-force hyperparameter grid search \cite{bergstra2012} to determine the optimal choice of the polynomial order, $p$, and the best regularized least squares solver, e.g., a lasso approach \cite{Friedman2010},  to solve for the polynomial coefficients, $\phi_{ij}$. 
For each of the k components, the surrogate with the lowest average root mean squared error across the five folds is selected for the optimization step. 
Our final surrogate model can then be written as 
\begin{linenomath}
\begin{align}%\label{eq:loss}
    \hat{f}(\bm\theta, \bm{\hat\phi}) &= \sum_{j=1}^k \hat{f}_j(\bm{\theta},\mb{\hat{\phi}}_j) \mb{\psi}_j,
\end{align}
\end{linenomath}
 resulting in a prediction $ \hat{f}(\bm\theta, \bm{\hat{\phi}}) \in \mathbb{R}^m$, where $\bm{\hat{\phi}}$ is the optimal choice of polynomial coefficients, comprising of the optimal choice of hyperparameters like the regularization and polynomial order.
We present the hyperparameter grid used for the PCE in \ref{app:hyper}.

In order to make this approach easily accessible
to non-ML experts, we provide a freely available and flexible Python machine learning library called \textit{tesuract} (\url{https://github.com/kennychowdhary/tesuract}) \cite{chowdhary_tesuract}. The library is compatible with the scikit-learn API \cite{scikit-learn}, which allows us to build on its vast library of ML tools for maximum flexibility. In fact, in addition to polynomial-based ML models, we can just as easily fit these components with random forests, Gaussian processes, and neural networks, all of which are 
accessed through the scikit-learn backend. Again, our philosophy is to let the data decide which model works best. Our choice of PCEs was based on 
experiments with our 250 ensemble runs which showed that, compared with neural networks, random forests, and Gaussian processes, PCEs performed the best (the full surrogate comparison results are presented in \ref{app:surrogate_comparison}). 

\subsection{Optimization}\label{sec:optimization_methods}

We will now describe the optimization step to learn 
%a set of ``optimized'' parameters 
$\mb{\hat\theta}$ using the surrogate with fitted parameters $\mb{\hat{\phi}}, $ learned via the hyperparameter cross validation approach described above.  
To construct the loss function, we specify a Gaussian likelihood for the observational fields centered at the fitted values of the surrogate model.
The principal consideration in our optimization scheme is the balance of magnitudes of the different variables and seasons in the loss function $L(\hat{f}(\mb{\theta}, \mb{\hat\phi}), \mb{Y}_{obs})$.

%To account for the differences in the magnitudes of errors across fields, we normalize each variable's RMSE by the spatial variance of the observational field. 

Let $Y_{p, \ell}$ denote the entry of $\mb{Y}_{obs}$ corresponding to the $p$-th target field and the $\ell^{th}$ grid point of that target field, and let $\hat{f}_{p, \ell}(\mb{\theta}, \mb{\hat{\phi}})$ be the surrogate prediction of $Y_{p, \ell}$ for parameters $\mb{\theta}$.
The normalized likelihood for this output can then be written as
\begin{linenomath}\begin{align}\frac{Y_{p, \ell}}{\sigma_{p}}  \overset{ind}{\sim} \mathcal{N}\left(\frac{\hat{f}_{p, \ell}(\mb{\theta}, \mb{\hat{\phi}})}{\sigma_{p}}, \frac{s_{p}^2}{w_{p, \ell}}\right)\label{eq:opt_norm}
\end{align}\end{linenomath} 
where $\{w_{p, \ell}\}_{\ell=1}^{m_p}$ are area weights on grid points for the $p$-th output field, and $\sigma_{p}^2 = \textrm{Var}_{\ell}(Y_{p, \ell})$ is the variance of the observational spatial field for the $p$-th output field.
In words, the distributional assumption in \eqref{eq:opt_norm} says that the standardized observations follow a normal (Gaussian) distribution with mean $\hat{f}_{p,\ell}(\mb{\theta}, \mb{\hat{\phi}})/\sigma_{p}$ and variance $w_{p, \ell}^{-1} s_{p}^2$ and that different locations and different variables are independent. 
%$\hat{f}(\mb{\theta}, \mb{\hat\phi})$ denotes the predicted values of proposed input parameters $\mb{\theta}$ for learned surrogate parameters $\mb{\hat\phi}.$
Here, $\sigma_{p}$ is the first step of normalization, approximately adjusting each field to the same scale. 
The terms $\{s_{p}^2\}$ are then the adjustments to this normalization which aim to more flexibly balance the fields. 
We emphasize that although $\{s_p\}$ is determined in the estimation process, one may also pre-specify $\{s_p\}$ to manually weight specific fields more than others.

We construct our Gaussian likelihood-based loss function through the joint log-likelihood over all target fields
\begin{linenomath}
\begin{align}
    \mathcal{L}\left(\mb{\theta}, \mb{s}^2, \mb{\hat{\phi}}, \mb{Y}_{obs}\right) 
    &\propto \sum_{p} \left(-\frac{e_{p}(\mb{\theta}, \mb{\hat{\phi}}, \mb{Y}_{obs}) }{ s_{p}^2}+  m_p \log\left(s_{p}^2\right) \right)
\label{eq:likelihood}\end{align}
\end{linenomath}where $\propto$ means ``proportional to,'' and $e_{p}(\cdot)$ is the weighted mean-squared error (MSE) for variable $p,$
\begin{linenomath}
\begin{align*}
    e_{p}(\mb{\theta}, \mb{\hat{\phi}}, \mb{Y}_{obs}) = \sum_{\ell=1}^{m_p} w_{p,\ell} \frac{\left(\hat{f}_{p,\ell}(\mb{\theta}, \mb{\hat{\phi}}) -  Y_{p, \ell}\right)^2}{\sigma_{p}^2}.
\end{align*}
\end{linenomath}

We highlight three options to find an optimized solution $\mb{\hat{\theta}}$: (1) maximum likelihood estimation (MLE), (2) maximum a priori estimation (MAP, a regularized approach to MLE), and (3) sampling from the posterior distribution of $\mb{\theta}$ via Markov Chain Monte Carlo (MCMC). We will discuss all three here but will only show results for the estimates obtained via the MAP and MCMC procedures. 
The MLE and MAP solutions are numerically found using the limited-memory optimization algorithm L-BFGS-B \cite{byrd1995limited}, which is commonly used to solve nonlinear optimization problems. 

The maximum-likelihood solution is
\begin{linenomath}
\begin{align*}
   \mb{\hat{\theta}}_{MLE}, \mb{\hat{s}}_{MLE}^2 &= \argmax_{\mb{\theta}, \mb{s}^2} \mathcal{L}\left(\mb{\theta}, \mb{s}^2, \mb{\hat{\phi}}, \mb{Y}_{obs} \right).
\end{align*}
\end{linenomath}
%This is obtained by effectively taking $ L(\hat{f}(\mb{\theta}, \mb{\hat{\phi}}), \mb{Y}_{obs}) = -\mathcal{L}(\mb{\theta}, \mb{s}^2, \mb{\hat{\phi}}, \mb{Y}_{obs})$ in equation \ref{eq:loss}. 

To obtain the MAP estimate, a prior distribution or regularization term is specified for $\mb{s}^2$.
Since our likelihood is defined for normalized values of $\bm{Y}_{obs}$, it is sensible to assign the same prior distribution to each element of $\mb{s}^2.$ This is equivalent to assuming there is no additional expert knowledge on whether some fields should be fit more closely to observations. %\note[BMW]{Good writing! The non mathematician understands!} \note[DY]{:celebration emoji:}
We specify the prior distribution $\mathcal{P}\left(s_{p}\right)$ as an inverse gamma distribution with parameters $\alpha$ and $\beta$, so that the logarithm of the joint posterior distribution can be written as 
\begin{linenomath}
\begin{align*}
    \log\left(\mathcal{P}\left(\mb{\theta}, \mb{s}^2 \middle| \mb{\phi},  \mb{Y}_{obs}\right)\right) 
    %&= C + g\left(\mb{\theta}, \mb{s}^2 | \mb{\phi},  \mb{Y}_{obs}\right),
    \propto  g\left(\mb{\theta}, \mb{s}^2 | \mb{\phi},  \mb{Y}_{obs}\right),
    \end{align*}
    %where $C$ is an unknown normalizing constant that does not affect the eventual optimization problem, 
    and 
    \begin{align*}
    g\left(\mb{\theta}, \mb{s}^2 \middle| \mb{\phi},  \mb{Y}_{obs}\right)&=\mathcal{L}\left(\mb{\theta},  \mb{s}^2, \mb{\phi},  \mb{Y}_{obs}\right) + \sum_{p}\log \left(\mathcal{P}\left(s_{p}\right)\right)  \\&= \mathcal{L}\left(\mb{\theta},  \mb{s}^2,\mb{\phi}, \mb{Y}_{obs}\right) + \sum_{p} (-\alpha - 1) \log\left(s_{p}^2\right) - \frac{\beta}{s_{p}^2}.
\end{align*}
\end{linenomath}
The hyperparameters $\alpha = 3$ and $\beta = 1/2$ are chosen.
The maximum a posteriori (MAP) estimate is then defined as
\begin{linenomath}
\begin{align*}
    \mb{\hat{\theta}}_{MAP}, \mb{\hat{s}}^2_{MAP} &=\argmax_{\mb{\theta}, \mb{s}^2} g\left(\mb{\theta}, \mb{s}^2 \middle| \mb{\hat{\phi}}, \mb{Y}_{obs}\right).
\end{align*}
\end{linenomath}

The inverse-gamma distribution is a common prior distribution used for the variance parameter of a normal distribution (for example, see section 2.6 of \cite{gelman1995bayesian}). 
The choices of the hyperparameters ensure that the prior distribution tends to have more mass for values between $0$ and $1$ compared to above $1$, which encourages lower values for these variances. 
These lower values then encourage the model output and observations to fit closer to each other rather than having a high tolerance for differences between them. %\note[BMW]{This is really about learning the tolerance for differences between model and obs for each target, right? Suggest clarifying that lower values mean less tolerance for model-data difference.} \note[DY]{I rewrote}%and therefore  parameter sets that lead to model output closer to the observations. 
The prior distribution for $\mb{\theta}$ is a uniform distribution across the parameter space, representing a lack of prior knowledge about its value. 
However, one may straightforwardly set a more informative prior distribution if desired.
More informative prior distributions on $\mb{s}^2$ can be used to place more weight on specific target fields if desired.
%\remove{As mentioned earlier, one can also choose $\mb{s}^2$ manually with ease in our implementation. }

Finally, one can use Bayesian Markov Chain Monte Carlo (MCMC) approaches to obtain samples from the posterior distribution 
\begin{linenomath}
    \begin{align*}
       \mathcal{P}\left(\mb{\theta}, \mb{s}^2 \middle|\mb{\hat{\phi}}, \mb{Y}_{obs}\right).
    \end{align*}
\end{linenomath}
These samples can then be used to provide a distribution of uncertainty around the MAP point, allowing one to visualize the relationships between the input parameters $\mb{\theta}$ and potentially consider additional areas of the parameter space such that the parameters fit the observations well. 

\section{Results}\label{sec:results}

In this section, we demonstrate the effectiveness of the autocalibration approach detailed in the previous sections on the E3SMv2 atmosphere model using the 250 PPE runs. 
We weigh the scalar value RESTOM equally in the optimization to one of the 32 latitude by longitude output fields. 
After removing missing data for latitude by pressure level fields (i.e.\ where the data would be underground), the total output size is $m=47{,}404$ points. %\note[BMW]{good to explain we dropped the missing data but not sure we need to say 47,404 points}\note[DY]{We should probably keep the total number so the reader knows the total output size. I reworded.}. 
We use a target RESTOM value of $0.70$ Watts per meter squared, 16 principal components (resulting in 86.6\% of variance explained of the original data), and 10-year simulations. 
The positive value of target RESTOM indicates that under perpetual 2010 forcing with prescribed present-day ocean surface temperature, we would expect more energy entering than exiting the top of the atmosphere due to the positive forcing caused by carbon dioxide and climate feedbacks. A relatively large range of RESTOM ($0.1$ to $1.5$ W/$\textrm{m}^2$) would be considered acceptable for this configuration because prescribed sea surface temperatures are an infinite source and sink of energy. %\note[DY]{Run by Benj}\note[BMW]{Nicely done but I edited it to my satisfaction. Suggest W/m2 instead of writing out watts per meter squared.} \note[DY]{Thank you :)}

We first present overall end-to-end results for the optimized parameters from our automated calibration approach. 
In table \ref{tab:auto_input_params}, we present the different solutions estimated by 
expert-tuned parameters for the released v2 model, which we refer to as ``v2 control" and denoted hereafter as $\mb\theta_{v2}$, the 
MAP estimate $\mb{\hat\theta}_{MAP},$ and the minimum and maximum bound considered by the PPE. 
The MAP estimates for \texttt{ice\_sed\_ai} and \texttt{clubb\_c1} are at the boundary of their parameter ranges. 
The other three parameters were estimated away from the boundary, and all parameters are substantially different compared to $\mb\theta_{v2}$. 

To evaluate performance, the E3SMv2 configuration with the optimized parameters $\mb{\hat{\theta}}_{MAP}$ was run for a 10-year simulation to obtain model output $f(\mb{\hat{\theta}}_{MAP})$. 
We then compare results with $f(\mb{\theta}_{v2})$.
Results from these output fields were compared against observations on a 180x360 grid for latitude/longitude fields and a 180x37 grid for latitude/pressure fields. Table \ref{tab:rmse_by_variable} displays the difference in root-mean-squared-error (RMSE) between the map estimate and the simulation run. The autocalibrated parameter set $\mb{\hat\theta}_{MAP}$ leads to a reduction in root-mean-squared-error (RMSE) for a majority of seasons and output variables compared to $\mb\theta_{v2}$, with a 2.7\% decrease on average across the spatial fields.
For many output fields, the improvements were substantial and in excess of a 10\% decrease in RMSE. 
These results were not uniform, however, and the automated-calibration parameter set did not provide improvements overall for DJF and for LWCF and PRECT. We note that clouds were the focus of a model hand-tuning effort in E3SMv1 \cite{ma2022}, and that the tuned parameter values mostly carried over into E3SMv2. Ma et al. also documented improvements in precipitation climatology, so these fields have less room for improvement by autocalibration. On the other hand, manually set values of $\mb{s}^2$ that focus on LWCF, SWCF, and PRECT could perform more comparably to the control simulation for these variables, potentially at the expense of sacrificing improvements in other variables. 

\begin{table}
\caption{Comparison of automated-calibration parameters with v2 control parameters and minimum and maximum values from which training data was sampled.}\label{tab:auto_input_params}

\centering
\begin{tabular}{lcccc}\hline
Input Parameter &   $\mb\theta_{v2}$  &  $\mb{\hat\theta}_{MAP}$&	 Minimum &    Maximum \\
\hline 
\texttt{ice\_sed\_ai} &500.00 & 1400.00 & 		350.00& 1400.00	\\ 
\texttt{clubb\_c1} & 2.40	& 1.00& 		1.00&5.00\\
\texttt{clubb\_gamma\_coef} &0.120	 & 0.312	&0.100 &0.500	\\ 
\texttt{zmconv\_tau}  & 3600.00	&4787.46& 	1800.00& 14400.00	\\ 
\texttt{zmconv\_dmpdz} & -0.00070	& -0.00042&	-0.00200& -0.00010	\\ \hline
\end{tabular}
\end{table}

 \begin{table}
 \caption{Comparison of autocalibrated and v2 control parameters: percentage change in root-mean-squared-error (RMSE) between time-averaged E3SMv2 output and observations. Green represents improvements when using the autocalibrated parameters. On average, the automated-calibrated configuration reduces RMSE by approximately 2.7 percent. 
 The autocalibration parameters also led to a RESTOM value of +0.47 under the F2010 configuration.}\label{tab:rmse_by_variable}
 \centering
 \begin{tabular}{lLLLLL}
Variable &	 \textrm{DJF} &    \textrm{MAM} &   \textrm{JJA} &  \textrm{SON} & \textrm{Avg.}\\
\hline 
 LWCF   &    \cellcolor{shadecolor2} 9.7 & \cellcolor{shadecolor} -1.3&  \cellcolor{shadecolor2}0.4&    \cellcolor{shadecolor2}10.0 & \cellcolor{shadecolor2}4.7  \\ 
PRECT &    \cellcolor{shadecolor2} 9.5 & \cellcolor{shadecolor2}4.1 &\cellcolor{shadecolor} -0.3 & \cellcolor{shadecolor2}11.8 &  \cellcolor{shadecolor2}6.3  \\ 
PSL      & \cellcolor{shadecolor2}4.3 &\cellcolor{shadecolor}-6.9 & \cellcolor{shadecolor}-5.3 & \cellcolor{shadecolor}-18.0 & \cellcolor{shadecolor}-8.6 \\  
 RELHUM & \cellcolor{shadecolor}  -1.7 &\cellcolor{shadecolor2}0.3 &\cellcolor{shadecolor2} 1.9 & \cellcolor{shadecolor2}0.4 & \cellcolor{shadecolor2}0.2 \\ 
 SWCF      &  \cellcolor{shadecolor2} 5.1  &\cellcolor{shadecolor} -0.3 &\cellcolor{shadecolor}-6.2& \cellcolor{shadecolor2} 2.0 & \cellcolor{shadecolor2}0.1  \\ 
 T        & \cellcolor{shadecolor}-0.3 & \cellcolor{shadecolor}-3.3 &  \cellcolor{shadecolor2}1.9 & \cellcolor{shadecolor}-4.0 & \cellcolor{shadecolor}-1.4 \\ 
 TREFHT  & \cellcolor{shadecolor}-7.2&\cellcolor{shadecolor} -10.0&  \cellcolor{shadecolor}-2.5 &\cellcolor{shadecolor} -10.3 & \cellcolor{shadecolor}-7.5  \\ 
 U        &\cellcolor{shadecolor2} 1.4   &\cellcolor{shadecolor} -10.6  &\cellcolor{shadecolor}-6.7 & \cellcolor{shadecolor}-10.8 & \cellcolor{shadecolor}-6.7 \\ 
 U200   &    \cellcolor{shadecolor2} 7.4& \cellcolor{shadecolor}-12.8& \cellcolor{shadecolor}-18.0  &\cellcolor{shadecolor}-7.3 & \cellcolor{shadecolor}-4.0 \\ 
U850     &  \cellcolor{shadecolor2} 5.7 &\cellcolor{shadecolor}-11.8 &\cellcolor{shadecolor}-16.1 &\cellcolor{shadecolor2}0.7 & \cellcolor{shadecolor}-5.4\\ 
Z500     &\cellcolor{shadecolor2} 4.0 &\cellcolor{shadecolor}-9.8 &\cellcolor{shadecolor} -7.1  &\cellcolor{shadecolor}-15.0 & \cellcolor{shadecolor}-2.7\\ \hline
Average & \cellcolor{shadecolor2}2.7 & \cellcolor{shadecolor}-5.7 & \cellcolor{shadecolor}-5.3& \cellcolor{shadecolor}-2.4 & \cellcolor{shadecolor}-2.7\\\hline
\end{tabular}
 \end{table}

\subsection{Surrogate Fit}

%First, we give results from the fitting process for the surrogate. 
Here, we further investigate the surrogate model fit. 
In table \ref{tab:selected_models}, we summarize the optimal polynomial order and coefficient fitting procedure, i.e., fit type, for the PCE surrogate selected by the cross-validation process for each principal component. 
In general, the first few principal components have a higher selected polynomial degree compared to the later principal components, suggesting that there are more complex relationships between the input parameters and the first few principal components. 
Using 64 computer cores on Chrysalis, selecting and fitting the surrogate in this configuration takes less than 45 seconds.

\begin{table}
    \centering
    \caption{Automatically-selected estimators for each principal component.} \label{tab:selected_models}
    \begin{tabular}{lllclll}
       PC & Fit type & Poly. order&& PC & Fit type & Poly. order \\ \hline  
       1 & linear & 8&& 9 & elastic net & 7 \\
       2 & lasso & 8 && 10 & lasso & 5 \\
       3 & lasso & 8 && 11 & lasso & 4  \\
       4 & lasso & 10 && 12 & lasso & 2  \\
       5 & lasso & 7 && 13 & elastic net & 5  \\
       6 & lasso & 8 && 14 & elastic net & 6  \\
       7 & elastic net & 5&& 15 & lasso & 8  \\
       8 & lasso & 5 && 16 & lasso & 2 \\ 
    \end{tabular}
\end{table}

\begin{figure}
    \centering
    \includegraphics[width = .98\textwidth]{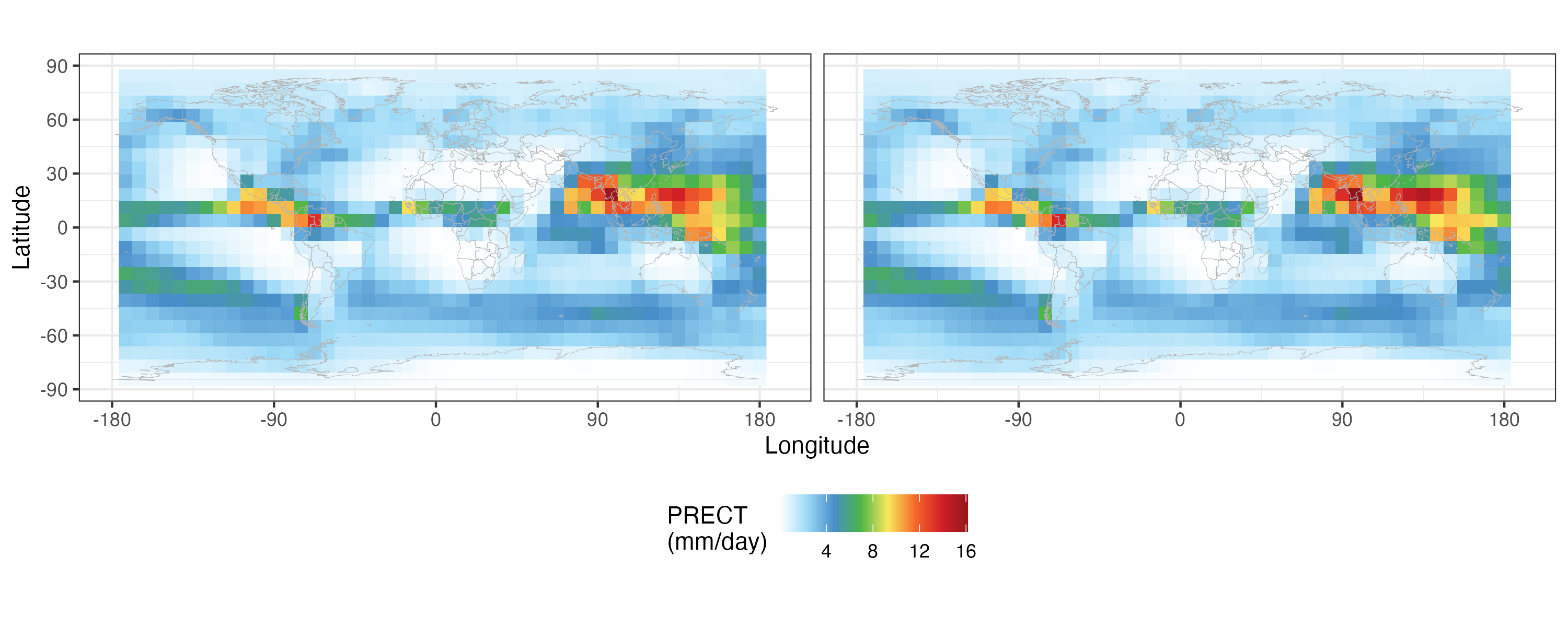}
    \includegraphics[width = .6\textwidth]{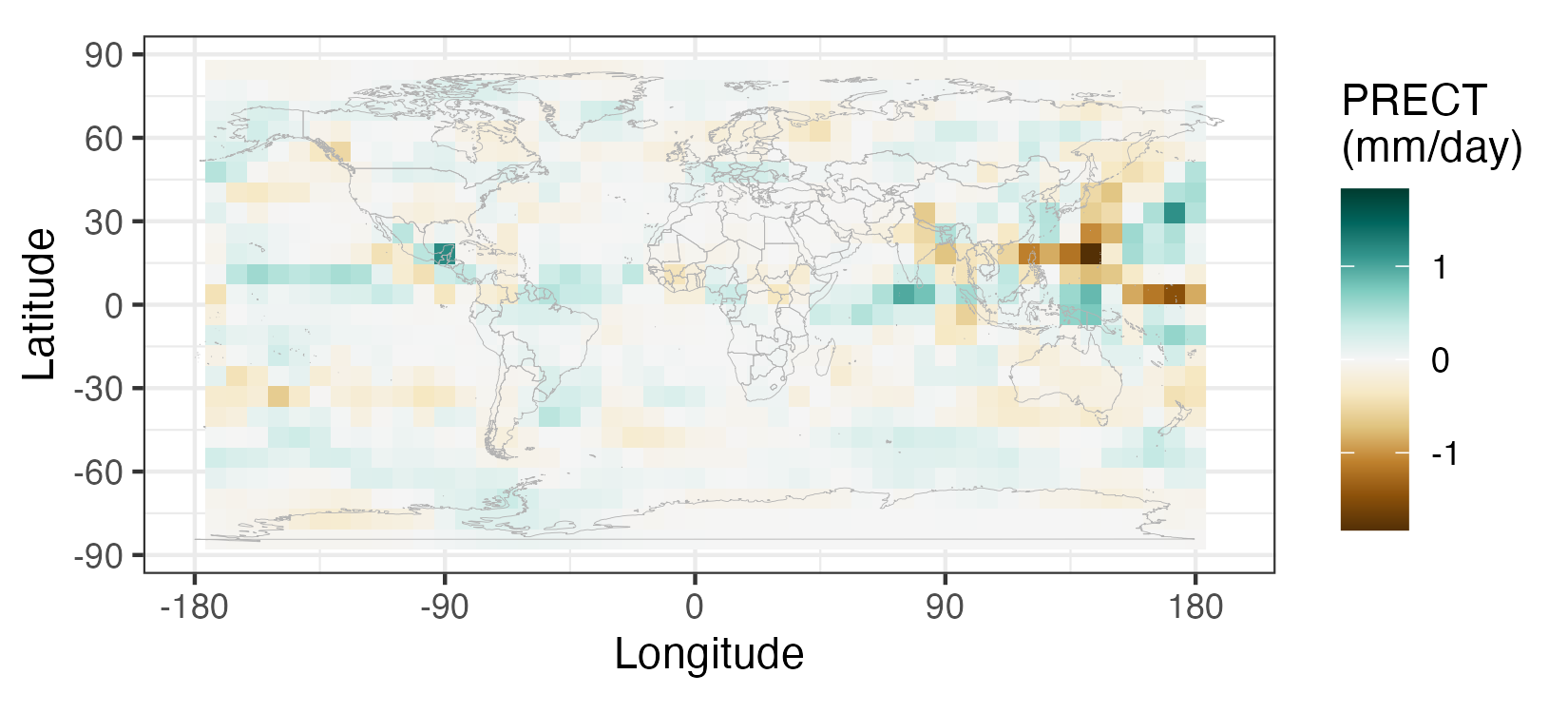}

    \caption{E3SM and surrogate-predicted output for time-averaged PRECT during JJA, plotted on a 24x48 grid. (Top Left) E3SM output, (Top Right) surrogate prediction, (Bottom) the difference between the E3SM output and the surrogate prediction. }%\note[DY]{let me know what you think of the updated plot}\note[LS]{i like it}}
    \label{fig:ex_prect_prediction}
\end{figure}

\begin{figure}
    \centering
        \includegraphics[width = .98\textwidth]{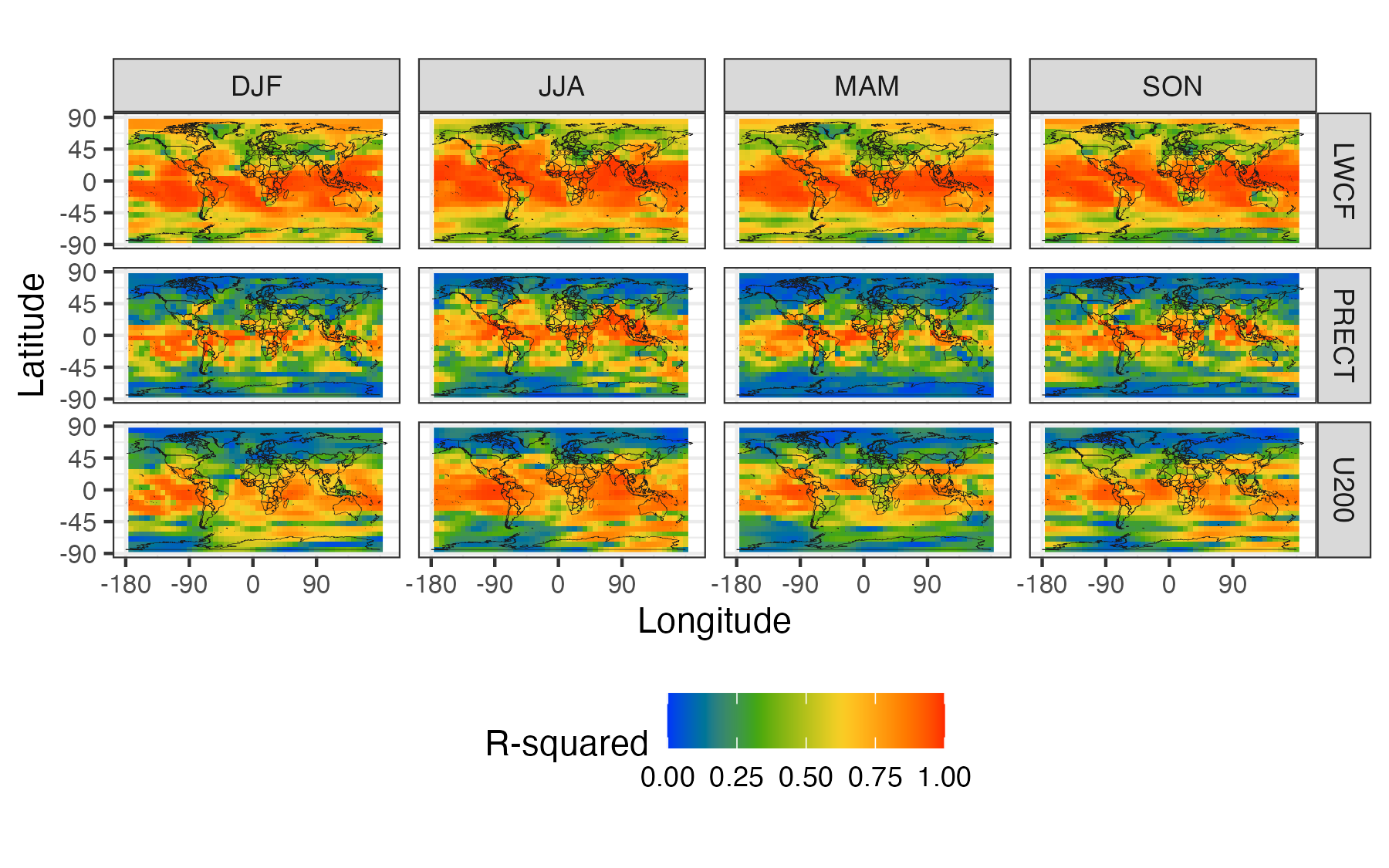}
        \includegraphics[width = .98\textwidth]{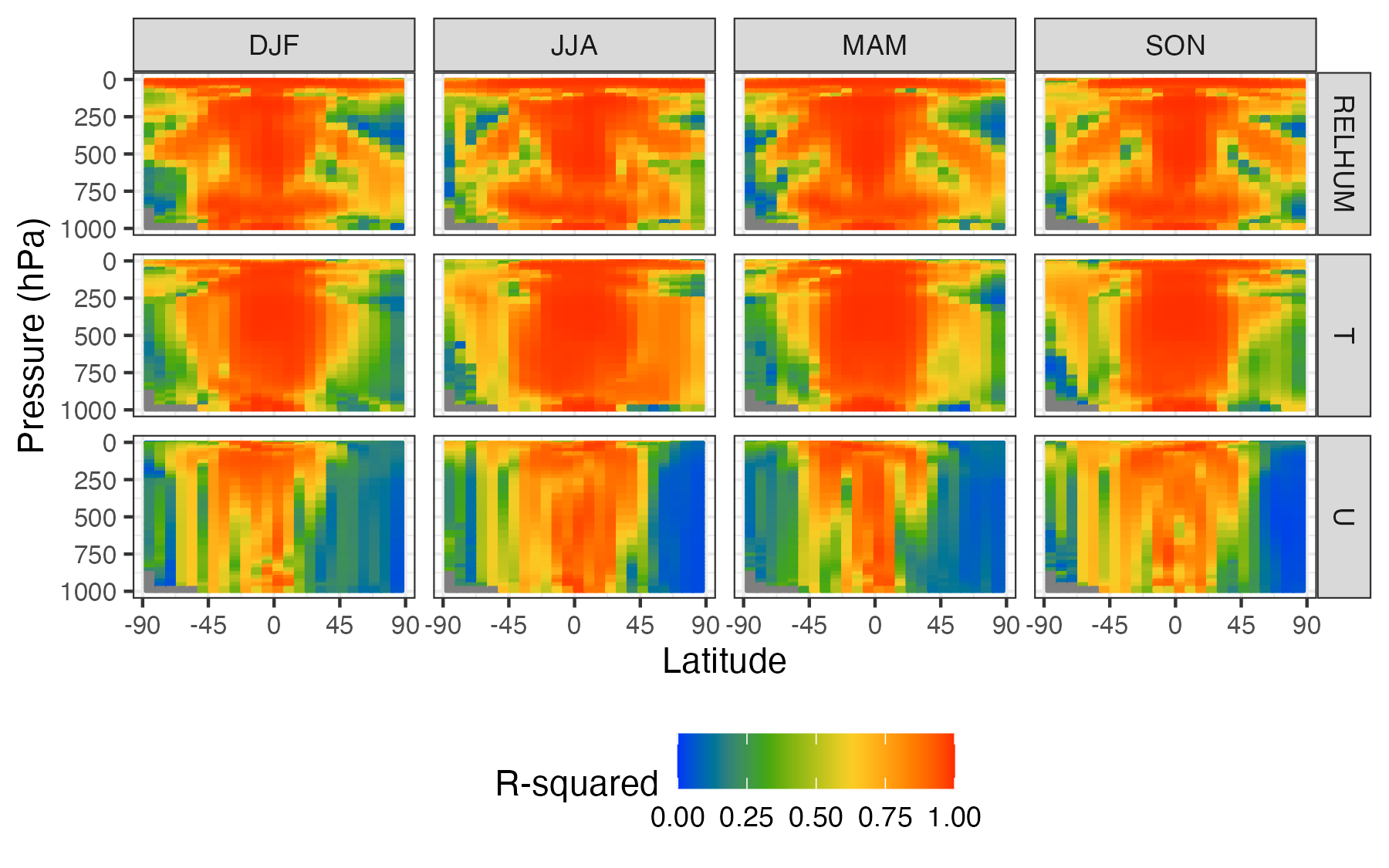}
    \caption{Surrogate R-squared ($R^2(\ell)$, proportion of explained variance) computed for different variables, seasons, and spatial locations. (Left) For 3 fields defined on longitude/latitude, (Right) for fields defined on latitude/pressure. The respective plots for all longitude/latitude variables are plotted in \ref{app:plots}.}% \note[DY]{Let me know what you think of the updated plots}\note[LS]{like these also}}
    \label{fig:rsquared_by_location}
\end{figure}

In figure \ref{fig:ex_prect_prediction}, we plot an example of a surrogate prediction from one output field, with the patterns of precipitation largely matching the E3SM output. 
Overall, the surrogate explains a substantial portion (R-squared of 0.48) of the variance in the perturbed parameter ensemble. 
%We note that using more simplified outputs (zonal or scalar targets) leads to a considerable increase in the reported R-squared. 
In addition to an overall R-squared value, we formulate a version of R-squared for each grid point. 
Formally, let $f(\mb{\theta}_i)_{\ell}$ be the $\ell$-th grid point for the $i$-th simulation run and $\hat{f}(\mb{\theta}_i, \mb{\hat{\phi}})_{\ell}$ its surrogate prediction. 
We evaluate 
\begin{linenomath}
$$R^2(\ell) = 1 - \frac{\sum_{i=1}^{n}\left(f(\mb{\theta}_i)_{\ell}- \hat{f}\left(\mb{\theta}_i,  \mb{\hat{\phi}}\right)_{\ell}\right)^2}{\sum_{i=1}^{n}\left(f(\mb{\theta}_i)_{\ell} - \overline{f}_\ell\right)^2}$$
\end{linenomath}
where $\overline{f}_\ell = \frac{1}{n}\sum_{i=1}^{n} f(\mb{\theta}_i)_{\ell}$ and plot $R^2(\ell)$ for a few variables in figure \ref{fig:rsquared_by_location}. 
The surrogate explains a higher proportion of variability in the tropics, suggesting that the surrogate focuses on high variability areas and that the output fields in these areas can be controlled using the five input parameters. 
For each of the output variables, we see similar surrogate performance across the four seasons in general.
The surrogate also predicts RESTOM very well with a high R-squared of 0.996. 

\begin{figure}
    \centering
    \includegraphics[width = .98\textwidth]{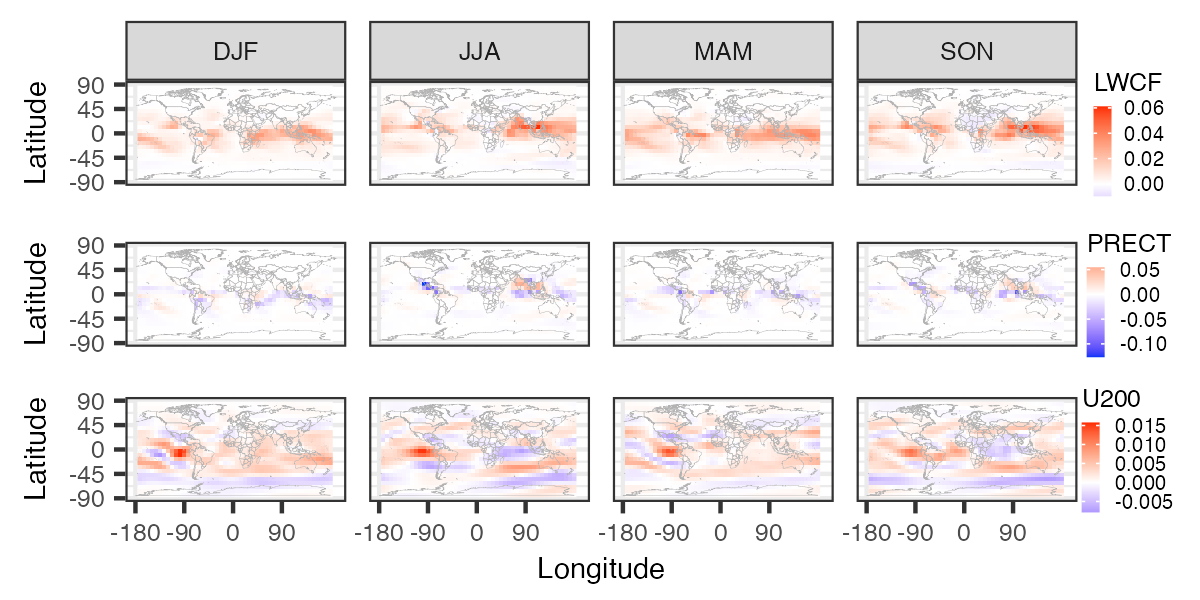}
    
    \includegraphics[width = .98\textwidth]{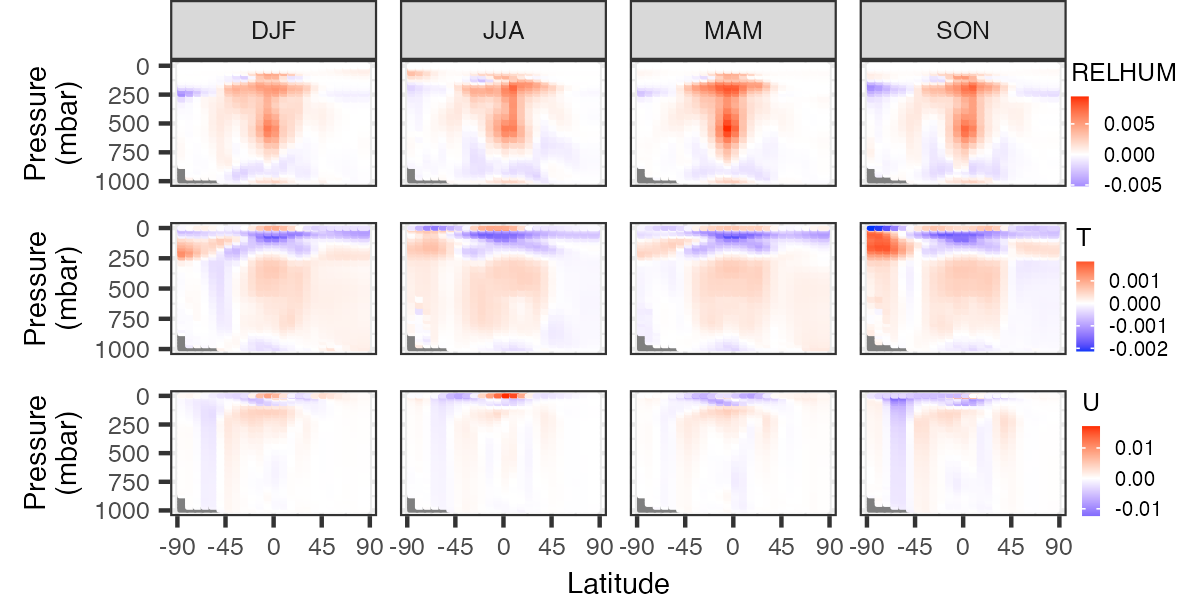}
    \caption{Values of the first principal component (PC) vector: (Top) for 3 fields defined on longitude/latitude and the first PC, (Bottom) for fields defined on latitude/pressure for the first PC. The respective plots for all longitude/latitude variables and RESTOM are plotted in \ref{app:plots}.}
    \label{fig:pc1_by_location}
\end{figure}

We can next investigate the ability of the principal components to represent the key spatial patterns and key modes of variability.
In figure \ref{fig:pc1_by_location}, we plot the first principal component vector. 
While further principal components may refine the covariance between different grid points, the first component describes the relationships between grid points along the largest mode of variability.
Based on this one-dimensional representation of the data, locations where the vector's absolute value is high generally have more variance compared to locations with values closer to $0$. 
Most fields thus appear to have more variability in the tropics rather than the polar regions. %\note[LS]{how do we know this? figure is very hard to read}
Also, two locations with the same color are likely positively correlated based on this first component. 
Therefore, we see that variables at the same location for different seasons are usually positively correlated. 
%\note[DY]{Leaving a general note to reevaluate our use of PCA, EOF, and ROM terminology}

\begin{figure}
    \centering
    \includegraphics[width = .5\textwidth]{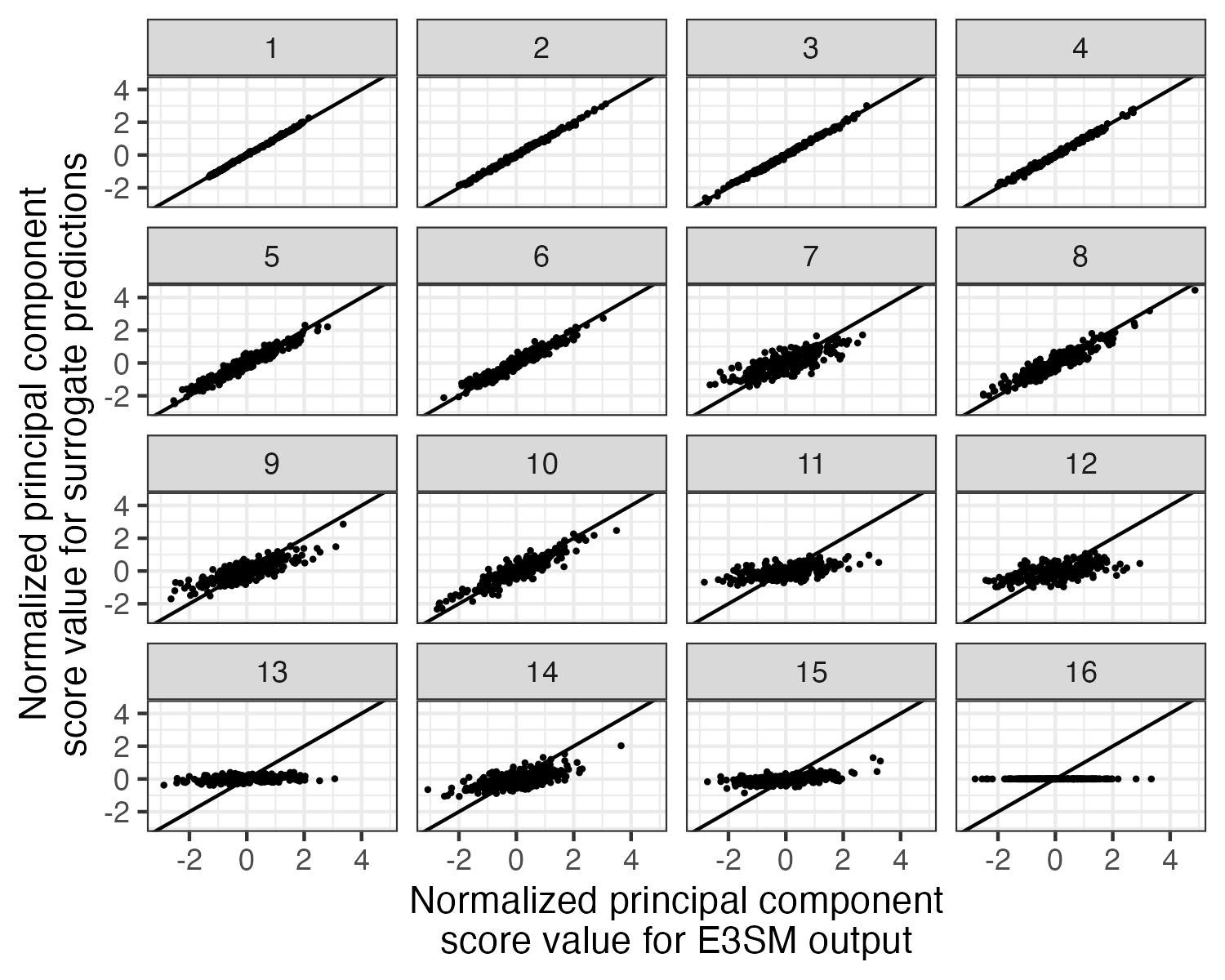}
    \includegraphics[width = .48\textwidth]{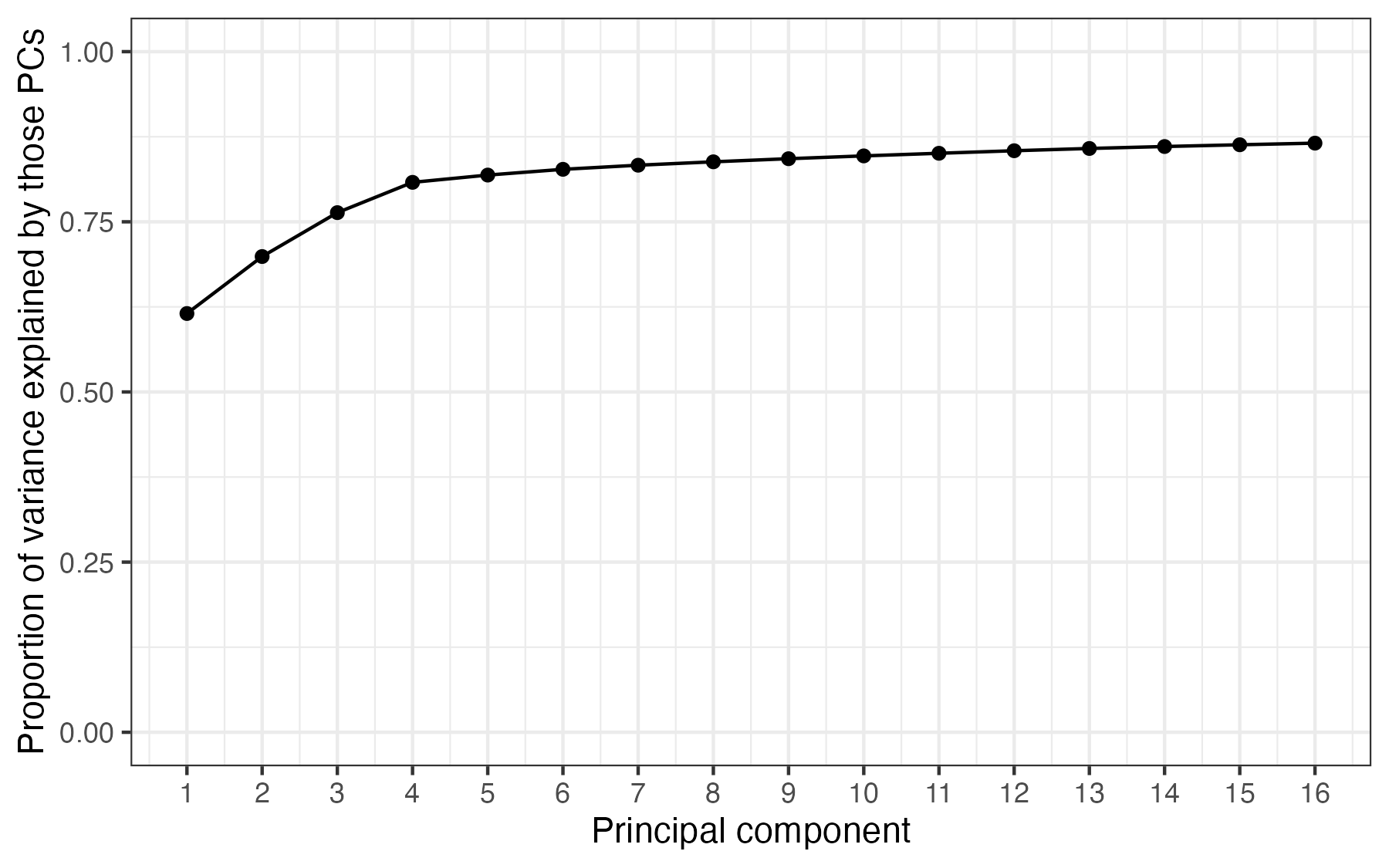}
    \caption{(Left) Principal component (PC) score/coordinate values for each principal component, comparing the simulation output and the surrogate predicted fields. The line y=x is plotted in each plot. (Right) The cumulative proportion of variance in the data explained by the first $k$ principal components, for $k=1$, $2$, $\dots$, $16$.  }
    \label{fig:pc_scatter}
\end{figure}

In figure \ref{fig:pc_scatter}, we compare the principal component scores for the E3SM simulation output and the surrogate's predictions. 
For each principal component score, we plot the 250 E3SM simulation runs, comparing to the line representing equal principal component scores for the simulation output and the surrogate predictions. 
The first principal components (for example, principal components 1 through 6) are predicted very well by the surrogate. 
These principal components are especially useful, in that the relationship between the input parameters and the principal components is especially strong. 
For later principal components, the surrogate is not able to predict as well, suggesting that these components are more noisy and are less explainable using the five input parameters. 

Another way to evaluate a principal component decomposition is using the proportion of variances the principal components explain. 
Figure \ref{fig:pc_scatter} shows the cumulative proportion of variance for 1-16 components.
Collectively, the first five principal components explain 81.9\% of variance and the first sixteen explain 86.6\%. 
We conclude that using the first sixteen principal components is suitable given its large proportion of the explained variance of the data. 
Since the later principal components explain far less of the variance in the data, the surrogate struggles with fitting these later components (as in the left panel of figure \ref{fig:pc_scatter}), and our experiments have shown that using more than 16 principal components does not necessarily improve the surrogate fit.

\subsection{Optimization and Validation}

Here, we describe the results of the optimization and validation procedures in greater detail.
Alongside the five automatically-calibrated parameters, normalization terms $\mb{s}^2$ for each of the 45 fields are optimized and are reported in table \ref{tab:opt_sigmas}. 
We note that smaller values mean that the optimization aims to fit these values more closely relative to other fields and weighs them more in the optimization problem. 
There are only minor differences in these values between seasons, yet there are substantial differences between variables. 
To some extent, the results from table \ref{tab:opt_sigmas} are in accordance with the RMSE results in table \ref{tab:rmse_by_variable}. 
For example, the results in table \ref{tab:opt_sigmas} suggest that fitting the LWCF, SWCF, and PRECT fields well is especially hard as they are weighed less in the optimization. 
In contrast, some of the variables where the automatically-calibrated parameters provide improvement (TREFHT, Z500) were weighed more in the optimization. 
This, again, suggests that if one uses $\mb{s}^2$ to weight the optimization of LWCF, SWCF, and PRECT more heavily, one would obtain improvements in these variables, possibly at the expense of performance in other variables.

 \begin{table}
 \caption{MAP estimates $\mb{\hat{s}}_{MAP}$ for each target field $p$ by variable and season. The $s_{p}$ for RESTOM had a value of $0.02$.}\label{tab:opt_sigmas}
 \centering
 \begin{tabular}{lcccc}
Variable &	 \textrm{DJF} &    \textrm{MAM} &   \textrm{JJA} &  \textrm{SON}\\
\hline 
 LWCF   &    0.48 & 0.42& 0.45 &    0.50 \\ 
PRECT &   0.49 & 0.43 & 0.58 & 0.53  \\ 
PSL      & 0.26 & 0.29 & 0.28 &  0.25  \\  
 RELHUM & 0.29 &0.26 & 0.36 & 0.26 \\ 
 SWCF      &  0.25  & 0.41 & 0.44 & 0.36   \\ 
 T        & 0.12 & 0.09 &  0.11 &0.09  \\ 
 TREFHT  & 0.06 & 0.06&  0.06 &0.07\\ 
 U        & 0.40  & 0.34  &0.23 & 0.33 \\ 
 U200   &    0.26 & 0.22& 0.19  & 0.22 \\ 
U850     &  0.26 & 0.24 &0.29 &0.24 \\ 
Z500     & 0.07 &0.08 & 0.08 &0.07\\ \hline
\end{tabular}

 \end{table}
 
\begin{figure}
    \centering
    \includegraphics[width = .95\textwidth]{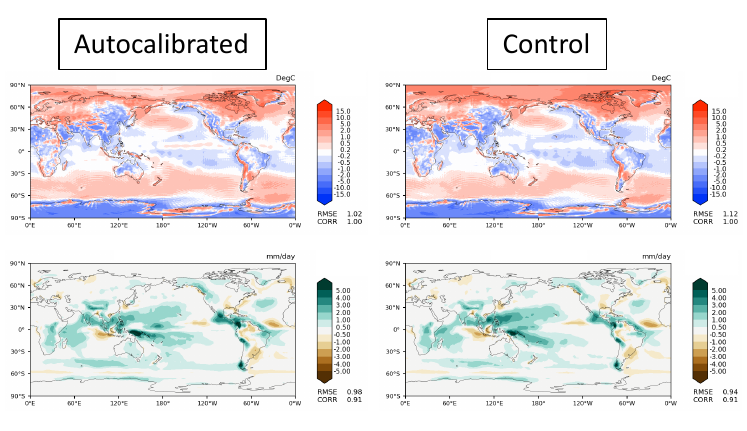}
    \caption{Simulated annual mean 2-meter temperature (top) and precipitation (bottom) minus observational targets for the auto-calibrated simulation (left) and the control simulation (right). Auto-calibrated temperature RMSE is improved, precipitation RMSE is degraded, and the large-scale spatial patterns of biases are consistent.} 
    \label{fig:diff_plot}
\end{figure}

While the RMSE comparisons in table \ref{tab:rmse_by_variable} provide field-level summary of performance of the automatically-calibrated parameter set, we give examples of the resulting spatial fields compared to observations for E3SM in figure \ref{fig:diff_plot}. 
Here, we compare annual means instead of seasonal means to summarize performance more succinctly. 
For the most part, the biases have the same spatial pattern in automatically-calibrated and control simulations. 
Some biases are structural to the model, and other biases may be sensitive to parameters that were not included in the PPE. 
In temperature (the top panels), the autocalibration reduces the RMSE and visually improves temperature differences in the northern high-latitudes.
For precipitation (the most degraded field, bottom panels), the autocalibration degraded RMSE. The spatial correlation in both fields was unchanged from the hand-tuned model. 
The surrogate also had more trouble predicting precipitation compared to other fields (see figures \ref{fig:rsquared_by_location} and \ref{fig:rsquared_by_location_all}), which then challenges the optimization effort. 
%The precipitat\note[BMW]{should we mention whether the surrogate prediction was wrong about precipitation? From this info, it is not clear whether the degraded precip was correctly predicted by the surrogate or a surprise when we ran E3SM}
%However, the automatically-calibrated parameter set provides some improvement, for example, in the tropical Pacific precipitation on the top panel of Figure \ref{fig:diff_plot}.

In addition to comparing the automatically-calibrated and control E3SM parameter sets to observations, we next aim to contextualize these results in the context of the perturbed parameter ensemble. 
One popular diagnostic tool to assess ESM performance is a Taylor diagram \cite{taylor_2001}. 
We give more details on its construction in \ref{app:taylor}.
The Taylor diagram visualizes the components of the centered mean-squared error between the model run and observations in one plot with polar coordinates: the standard deviation of the model field normalized by the standard deviation of the observational field is the radius, and correlation between the two fields is the angular component of the polar coordinates. 
As a result, the distance between the evaluated point and the idealized point at $(1,0)$, representing a CMSE of $0$, is the square-root of the normalized CMSE. 

\begin{figure}
    \centering
    \includegraphics[width = .48\textwidth]{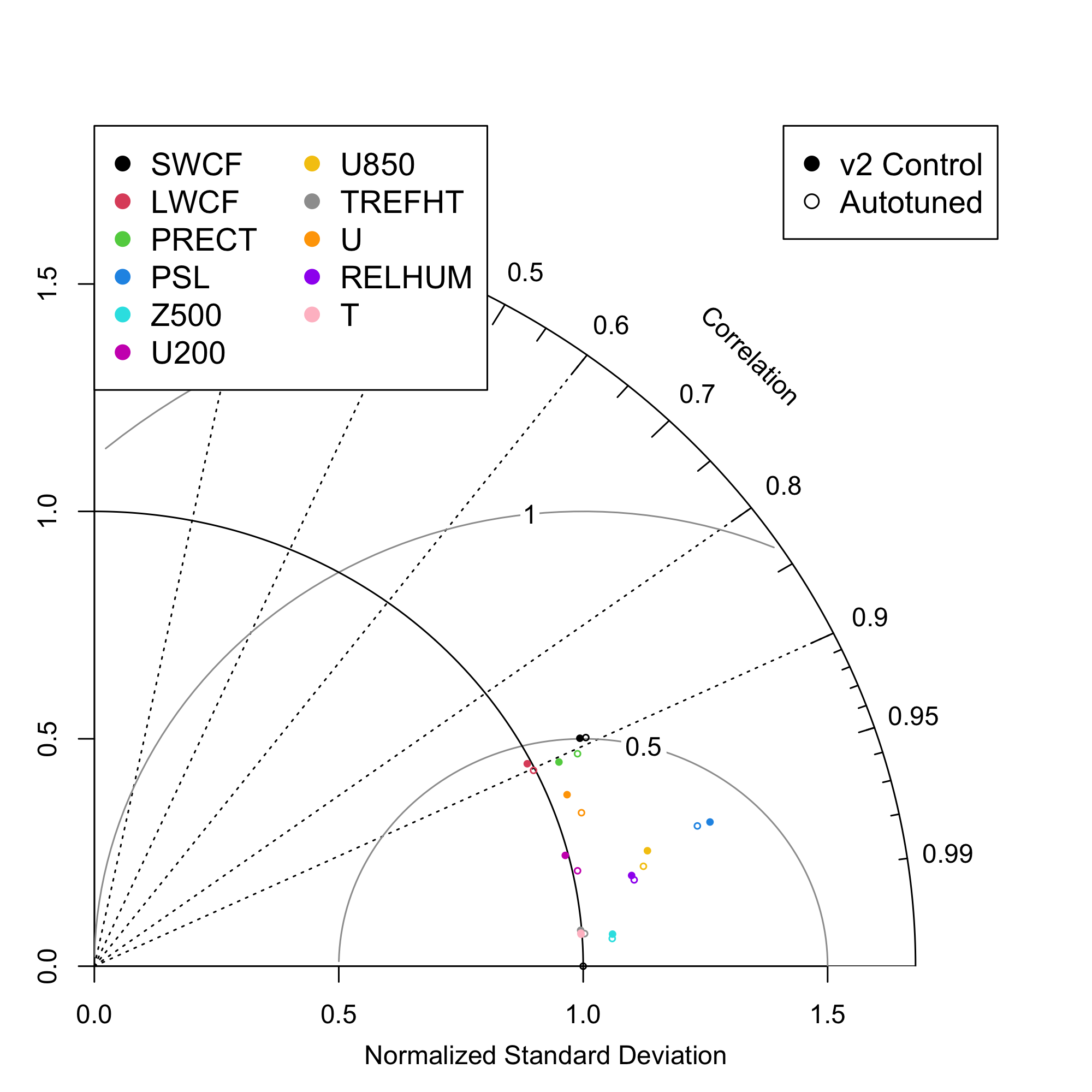}
    \includegraphics[width = .48\textwidth]{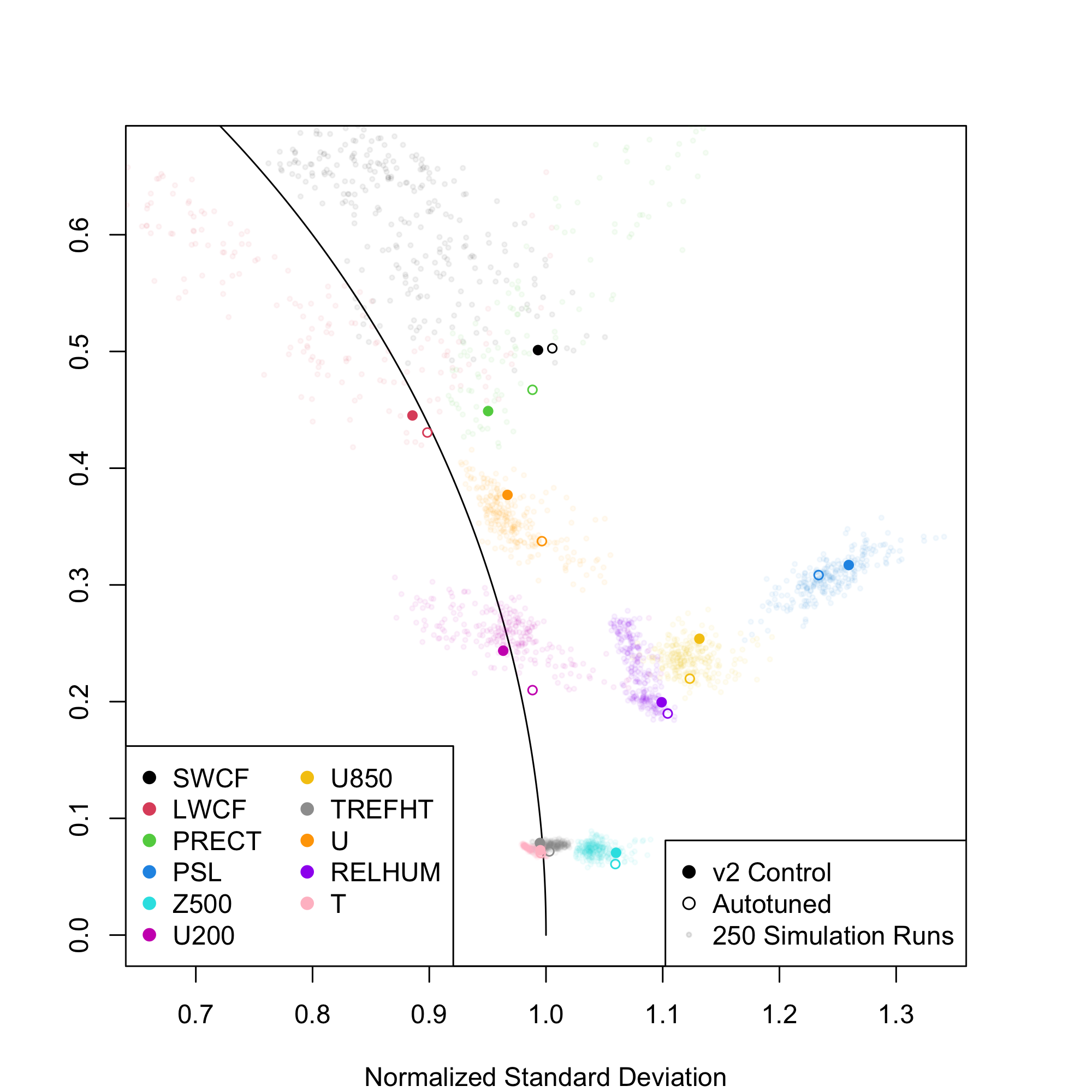}
    \caption{Taylor diagram of automatically-calibrated and control parameters for the season MAM. (Left) The entire Taylor diagram for the v2 control and automatically-calibrated parameter set. (Right) A zoomed-in version of the plot on the left, with the performance of the 250 simulation runs also plotted lightly in the background.}
    \label{fig:taylor}
\end{figure}

In figure \ref{fig:taylor}, we plot area-weighted versions of these statistics in a Taylor diagram for a single season (March-April-May, or MAM) for each of the variables on the same plot. 
On the left of figure \ref{fig:taylor}, the diagram is plotted in totality comparing automatically-calibrated and control parameter sets to the observational fields. 
Points plotted directly on the circle centered at $(0,0)$ with radius $1$ indicate that the model and the observations have the same standard deviation for that variable.
The circles centered at $(1, 0)$ represent contours of equal values of the normalized $\sqrt{\textrm{CMSE}}$, so that, in general, points closer to $(1,0)$ represent models with lower centered mean-squared error. 
We see that, in general, the autocalibrated parameters provide modest improvements compared to the control set of parameters for most variables. 
The Taylor Diagram is an important cross check, because autocalibration targets RMSE instead of centered RMSE (CRMSE). Therefore, the optimal autocalibrated simulation could reduce bias at the expense of spatial correlation. However, examination of the Taylor Diagram shows that improvements in RMSE generally do not come at the expense of spatial correlation.

The right-most plot in figure \ref{fig:taylor} shows a close up view of the same points along with the performance of each of the 250 PPE runs. 
For each variable, one should not expect the autocalibrated parameters to perform substantially better than the best of the 250 of the PPE runs. 
Therefore, for some variables like U200, RELHUM, and LWCF, we would not expect there to be much improvement in these variables for this season without using a different PPE. 
Conversely, it may be possible to improve the results for other variables like PSL by weighing them more in the optimization since some of the 250 PPE runs perform better in terms of CMSE. 
Since we are interested in the model performance for all fields, we would not expect that the automatically-calibrated parameter set perform better than all 250 PPE runs for all fields or even any field, yet for most variables the automatically-calibrated parameter set is competitive with or better than the top 25\% of simulations in terms of CMSE.

The Taylor diagram also informs us on how the PPE was designed. 
For the temperature variables T and TREFHT, the 250 simulation runs are tightly grouped, suggesting that there isn't much variability in the PPE associated with these variables. 
If larger variance and flexibility is desired for these variables, a future PPE should be designed accordingly. 
For other variables like LWCF, SWCF, and PRECT, we see that the PPE induces a larger range of outcomes, reflecting that the automatic calibration approach has the ability to choose parameters that affect these variables substantially. 
Overall, this plot emphasizes that the PPE enables the autocalibration approach to provide substantial improvements, yet overall biases in the model cannot be entirely overcome without modifying or expanding the PPE. 

\subsection{Bayesian Implementation}\label{sec:bayes}

In addition to providing a maximum a posteriori (MAP) estimate of the input parameter set, one may visualize the objective function to better understand uncertainty associated with the predicted parameters. 
In Bayesian statistics, Markov Chain Monte Carlo (MCMC) iteratively samples from the objective function to provide a distribution of the relevant input parameters. 
A Bayesian approach would allow the identification of a multimodal posterior, pointing to ranked solutions.
We run 200 independent chains of MCMC to avoid sensitivities to the algorithm initialization.
Of these chains, one half are initialized as random samples from the 250 input parameters based on Latin Hypercube Sampling; the remaining half are chosen uniformly within the set parameter bounds.
For each chain, we iteratively obtain 8,000 samples. 
The first 5,000 are discarded as ``burn-in,'' and every tenth sample from the remaining 3,000 samples is retained in a ``thinning'' step to decrease the correlation between neighboring samples. 
This results in 60,000 total samples from the combined 200 chains.

\begin{figure}
    \centering
    \includegraphics[width = .48\textwidth]{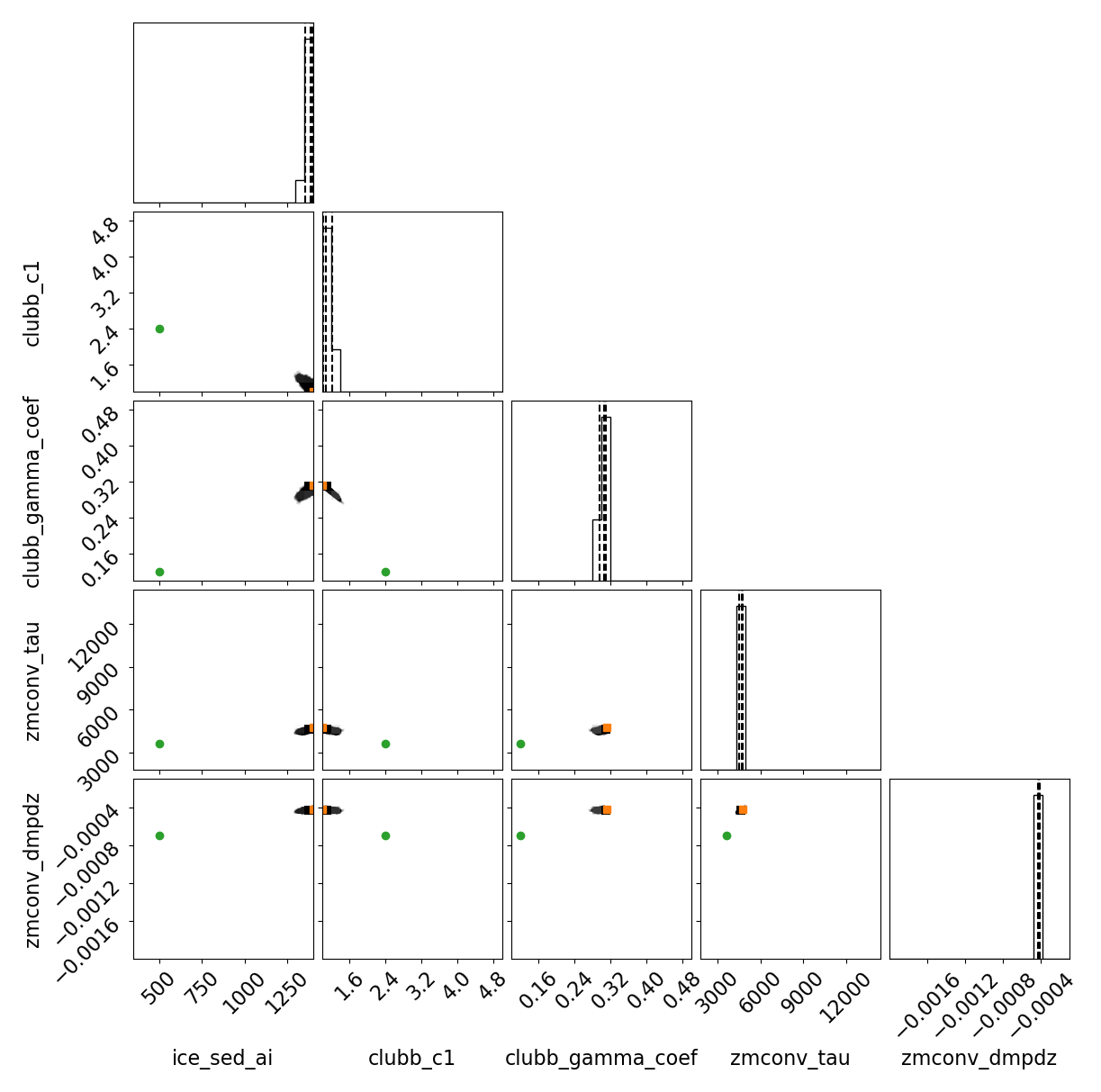}
    \includegraphics[width = .48\textwidth]{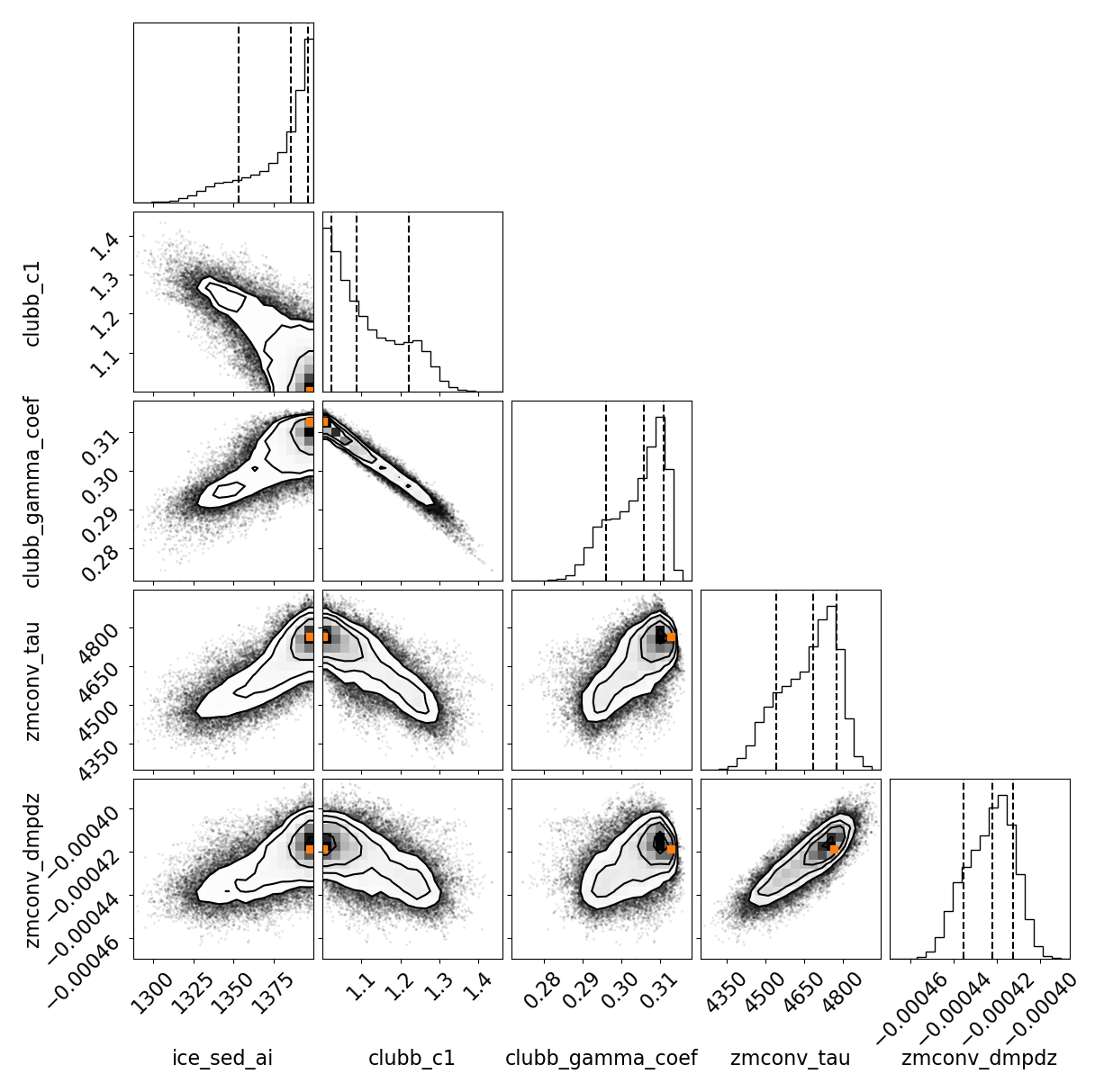}

    \caption{MCMC results: (Left) with plot bounds specifying the range of the parameter bounds searched; (Right) zoomed in to the bounds of MCMC samples.
    The vertical lines on the histograms represent the 16\%, 50\%, and 84\% quantiles of the distribution of each input parameter. 
    The orange squares represent the MAP points, while the green circles represent the E3SMv2 default parameter set. %\note[BMW]{Probably keep both plots, even though I think it's showing our over-confidence. It also shows we started with a wide range}
    }
    \label{fig:mcmc}
\end{figure}

In figure \ref{fig:mcmc}, we plot the results for the optimized parameter selection. 
Across the diagonal of each plot are histograms of the MCMC samples for each parameter.
On the off-diagonal of each plot are scatter and density plots of the samples describing the relationship between each pair of input parameters. 
The posterior distributions suggest bimodality with respect to the clubb parameters and ice\_sed\_ai. 
This bimodality indicates there are two combinations of input parameters that give output fields matching well to observations: one with a value of \texttt{clubb\_c1} on the lower boundary, a value of \texttt{ice\_sed\_ai} on the upper boundary, and a moderate value of \texttt{clubb\_gamma\_coef}; and one with a slightly higher value of \texttt{clubb\_c1}, a slightly lower value of \texttt{ice\_sed\_ai}, and a slightly higher value of \texttt{clubb\_gamma\_coef}.
In addition, due to the positive relationship between \texttt{zmconv\_tau} and \texttt{zmconv\_dmpdz} in the plot, one can retain fidelity with observations (to some extent) by increasing or decreasing both parameters simultaneously nearby the MAP point.
We see that $\mb\theta_{v2}$ are substantially different from $\mb{\hat{\theta}}_{MAP}$, and the distribution of the parameters are quite localized compared to the large range considered in the input parameters. These narrow posterior distributions indicate we are likely underestimating the uncertainty represented in these posterior distributions since they do not reflect variability due to surrogate model fit. As implemented here, figure \ref{fig:mcmc} represents the posterior distribution around the MAP estimate $\mb{\hat\theta}_{MAP}$.

\section{Discussion} \label{sec:disc}

In this work, we introduce a practical, fast, and flexible approach for automatic calibration using an ML-based surrogate model for spatially varying fields and present its results on the E3SMv2 atmosphere model. In this setting, we show the effectiveness of the automatic calibration framework, improving root-mean-squared-error on output fields compared to an expert-tuned control set of parameters by 2.7\% on average. 
This application shows that the surrogate flexibly fits the output fields, and the optimization identifies suitable parameters. 
Furthermore, through Bayesian estimation, we can provide a distribution of the parameters, incorporating some uncertainty into the estimates and potentially providing ranked solutions. 
This distribution also shows how the different input variables interact to give parameter sets in accordance with the optimization problem. 
As alluded to above, multiple solutions can be found quickly using this approach through user-chosen weights ($s_p^2$) of the target fields.

Our general automatic calibration framework has a number of advantageous qualities.
First, through empirical orthogonal functions (EOF), we can decompose the full ensemble of model output into a reduced space of orthogonal components, each of which represents a unique mode of variability.
The framework then seamlessly takes advantage of a large number of output fields.
In addition, we give flexibility in each step of the process, empowering the user to choose the number of principal components, $k$, output fields of interest, choice of surrogate, loss function and field-specific weights in the optimization function.  
Surrogate model parameters are learned automatically through cross-validated hyperparameter tuning.

Although the combination of PCE and a Gaussian loss function worked well in our case compared to other popular and successful surrogate models, other surrogates and loss functions can be streamlined into our approach. 
In particular, while the preferred candidate for our ML-based surrogate is polynomial-based, we also show that a well tuned GP, random forest, and/or fully connected neural network, can achieve similar levels of accuracy, albeit slightly worse than a well-tuned PCE. Our philosophy is to let the data decide which model is best in terms of a performance metric (e.g.\ cross-validated R-squared), and, while PCE may work better for our data, perhaps GPs or neural networks may work better for a different data set. %The framework we propose allows for this flexibility.

There are a number of opportunities to improve upon this work. 
Future work could include estimating $\bm{\hat\theta}$ within a full Bayesian framework. 
This would include the uncertainty due to the surrogate model fit through assigning a prior distribution to $\mb{\phi}.$ This is not a straightforward procedure for regularized PCE surrogates although it is an active area of research \cite{lu2015limitations}. Alternatively, to achieve this, one could specify a surrogate that is more adaptable to the Bayesian framework such as Gaussian process approaches. We are currently pursuing the feasibility of using Bayesian Adaptive Smoothing Splines \citep[BASS, ][]{yue2014bayesian} or Bayesian Additive Regression Trees \citep[BART, ][]{chipman2010bart} which have shown promise for automated calibration in other applications and are already suited for Bayesian estimation.

We also recognize that the assumption of independence between different spatial locations and between different variables in the optimization step is a computational approximation and not perfect theoretically. 
When using full spatial fields, this results in an overestimate of the amount of independent data used in the optimization problem. 
Specifying dependence, however, between a large number of locations for multivariate output fields is challenging both conceptually and computationally, and the independence assumption implies simpler computation and a first-order comparison to the observations.
We show that the optimized parameters from this approach work very well. An alternative solution would be to define the likelihood in the reduced space rather than the observed space, which we are currently exploring.

Although we have a large target space, the input space of five parameters in our exemplar was relatively small. We are currently working towards using this approach to tune E3SMv3 and test the robustness of our method with 14 specified atmospheric (input) parameters. This will require a larger Latin Hypercube sample to ensure high-fidelity of our surrogate. 
Whereas E3SMv2 default parameters had been set before our analysis, effectively providing a base case for comparison, tuning E3SMv3 will require extending and adjusting our approach as we interact with model tuning experts. 
For example, we may elicit subject matter expertise to adjust values of $\mb{s}^2$ based on priorities on the different output fields, and through these adjustments, we expect to provide multiple ``optimal" parameter sets with different climate sensitivities.
%Lastly, it is worth noting that although the ultimate target for automatic calibration, adapting our approach for the fully-coupled E3SMv3 model would be a more challenging endeavor. 
Lastly, it is worth noting that although the ultimate target for model tuning, the fully-coupled E3SM model, including an active ocean model, would be a more challenging computational endeavor.

\clearpage
\appendix

\section{Details on PCE and Hyperparameter Grid}\label{app:hyper}

Briefly describing the process to estimate $\mb{\phi}$, we use penalized optimization. 
Consider the minimization problem
\begin{linenomath} \begin{align*}
    \min_{\mb{\phi}_j}\sum_{i=1}^n \left(\eta_{ij} - \hat{f}_j(\mb{X}_{i\cdot}, \mb{\phi}_j)\right)^2+\lambda \textrm{Pen}(\mb{\phi}_j),
\end{align*}\end{linenomath}where $\eta_{ij}$ is the $i$-th entry of $\mb{\eta}_j$ and $\hat{f}_j(\mb{X}_{i\cdot}, \mb{\phi}_j)$ is a polynomial of the inputs $\mb{X}_{i\cdot}$ with order (or ``degree'') $D$ with coefficients specified by $\mb{\phi}_j$. 
The penalty $\lambda \textrm{Pen}(\mb{\phi}_j)$ is a regularization (e.g.\ lasso or elastic net) term to avoid overfitting of $\hat{f}_j$ to the training data, with its size determined by $\lambda>0$. 
The optimization problem may also be written 
\begin{linenomath} \begin{align*}
    \min_{\mb{\phi}_j}\sum_{i=1}^n \left(\eta_{ij} - \sum_{i=0}^P \phi_{ij}L_i(\bm{\theta})\right)^2+\lambda \textrm{Pen}(\mb{\phi}_j).
\end{align*}\end{linenomath}

For the polynomial chaos expansion, we consider polynomials of order 1 through 12. 
The type of interactions between input variables was chosen between ``total order'' (all possible interactions up to a prescribed polynomial order) and ``hyperbolic'' (a reduced set of interactions); see \cite{chowdhary_tesuract} for more information. 
The fit type was chosen between linear, elastic net, or lasso. 
The penalty parameters for elastic net and lasso were chosen from 20 values evenly-spaced on the log scale between $10^{-8}$ and $10^4$.  

\section{Comparison of Surrogate Models}\label{app:surrogate_comparison}

We compare the polynomial chaos expansion to alternative choices of surrogate model using cross-validation scores (Table \ref{tab:surrogate_comparison}).
Random forests are trained with varying number of trees, number of selected variables per tree, and tree depth.
These hyperparameters are chosen by cross-validation similar to the approach for the polynomial chaos expansion.
A Gaussian process (GP) regression model is also considered, with either a radial basis function or a Mat\'ern covariance chosen by cross-validation.
Finally, we consider a multilayer perceptron with varied number of layers and neurons in each layer, varying learning rate, and varying solver chosen by cross-validation.
For each of the four models, we run the automatically-calibrated process including surrogate fitting, surrogate cross-validation, and optimization of parameters. 
We compare total running time and cross-validated surrogate performance in table \ref{tab:surrogate_comparison}.
The polynomial chaos expansion has the best performance and also runs substantially faster than the random forest and the multilayer perceptron. 
We also compare the optimized parameters in table \ref{tab:surrogate_params_comparison}. 
Overall, the surrogates lead to similar values of the optimized parameters.
One exception is the random forest model, which has substantially different parameters for all parameters except for \texttt{ice\_sed\_ai}.
While choice of surrogate may depend on the particular application of the automatic calibration approach, we conclude that the polynomial chaos expansion performs as well or better than other state-of-the-art surrogate models. 
\begin{table}
    \centering
    \caption{Comparison of surrogate models using cross-validated scores: polynomial chaos expansion (PCE), random forest (RF), Gaussian process regression (GPR), and multilayer perceptron (MLP). For each surrogate model, we run on a single node in Chrysalis. } \label{tab:surrogate_comparison}
    \begin{tabular}{lllll}
        \hline
       Name & R-squared & RMSE & Median absolute error &  Time (mm:ss)\\ \hline 
       PCE & 0.478 & 7.36 & 5.04 &  07:42  \\
       RF &0.444  & 7.60 & 5.19 &  45:01   \\
       GPR &0.468 & 7.45 & 5.08& 06:52  \\
       MLP &0.466 & 7.46 &  5.10 &42:45  \\
    \end{tabular}
\end{table}

\begin{table}
    \centering
    \caption{Comparison of surrogate models by optimized parameters: polynomial chaos expansion (PCE), random forest (RF), Gaussian process regression (GPR), and multilayer perceptron (MLP).} \label{tab:surrogate_params_comparison}
    \begin{tabular}{llllll}
        \hline
       Name & \texttt{ice\_sed\_ai} & \texttt{clubb\_c1} & \texttt{clubb\_gamma\_coef} & \texttt{zmconv\_tau} & \texttt{zmconv\_dmpdz}  \\ \hline  
      PCE & 1400.00 & 1.00 & 0.312 &4787.46 & -0.00042  \\
      RF  & 1326.58& 1.90 & 0.259 & 6078.88 & -0.00013\\
       GPR &1106.87& 1.00 &  0.322 &3565.75 & -0.00045\\
       MLP  & 1006.56 & 1.00 & 0.329 & 3668.53 & -0.00049\\
    \end{tabular}
\end{table}

\section{Description of Taylor Diagram}\label{app:taylor}
We more concretely describe the Taylor diagram of \cite{taylor_2001}. 
Let $\mb{Y}_{obs} = [Y_{obs,\ell}]_{\ell=1}^{m}$ denote a vector of observations and $f(\mb{\theta}) = [f_\ell (\mb{\theta})]_{\ell=1}^m$ be a corresponding vector of model output at parameters $\mb{\theta}$.
If $\overline{Y}_{obs}=  m^{-1} \sum_{\ell=1}^m Y_{obs,\ell}$ and $\overline{f}(\mb{\theta}) = m^{-1} \sum_{\ell=1}^m f_\ell(\mb{\theta})$ are the sample averages and $\sigma_{Y}^2 =\frac{1}{m} \sum_{\ell=1}^n (Y_{obs,\ell} - \overline{Y}_{obs})^2$ and $\sigma^2_{f(\mb{\theta})} = \frac{1}{m} \sum_{\ell=1}^m (f_\ell(\mb{\theta})- \overline{f}(\mb{\theta}))^2$ are the sample variances, then the centered mean-squared error is 
\begin{linenomath}
\begin{align*}
    \textrm{CMSE}(\mb{\theta}) &= \frac{1}{m} \sum_{\ell=1}^m \left[(Y_{obs,\ell} - \overline{Y}_{obs}) - (f_\ell(\mb{\theta})- \overline{f}(\mb{\theta}))\right]^2= 
       \sigma^2_Y + \sigma^2_{f(\mb{\theta})} - 2\sigma_X\sigma_{f(\mb{\theta})} \rho_{Y,f(\mb{\theta})}
\end{align*}where \begin{align*}
    \rho_{Y,f(\mb{\theta})} =\frac{\frac{1}{m} \sum_{\ell=1}^m (Y_{obs,\ell} - \overline{Y}_{obs}) (f_\ell(\mb{\theta})- \overline{f}(\mb{\theta}))}{\sigma_Y\sigma_{f(\mb{\theta})}}
\end{align*}is a sample correlation coefficient between the two fields. 
Note that bias-corrected versions of the variance and correlation estimates can be used with $(m-1)^{-1}$ replacing $m^{-1}$. 
The Taylor diagram visualizes the components of the centered mean-squared error in one plot with polar coordinates: the normalized standard deviation $\sigma_{f(\mb{\theta})}/\sigma_Y$ is the radius, and $\cos^{-1}(\rho_{Y,f(\mb{\theta})})$ is the angular component of the polar coordinates. 
As a result, the distance on the plot between (a) the point in polar coordinate of the comparison \begin{align*}
    (\sigma_{f(\mb{\theta})}/\sigma_Y, \cos^{-1}(\rho_{Y,f(\mb{\theta})})) 
\end{align*} and (b) the idealized point representing CMSE $(\mb{\theta})=0$ 
\begin{align*}
    (\sigma_Y/\sigma_Y = 1, \cos^{-1}(1) = 0)
\end{align*} is the square-root of the normalized $\textrm{CMSE}(\mb{\theta})$: \begin{align*}
   \sqrt{\frac{\textrm{CMSE}(\mb{\theta})}{\sigma_Y^2}}.
\end{align*}
In other words, the performance of a model in terms of $\textrm{CMSE}(\mb{\theta})$, standard deviation $\sigma_{f(\mb{\theta})}$, and correlation with the observations $\rho_{Y,f(\mb{\theta})}$ can be visualized in the same diagram. 
While this ignores bias $\overline{Y}_{obs} - \overline{f}(\mb{\theta})$ between the fields, it provides a comprehensive look at the second-order error between the two fields. 
\end{linenomath}
\section{Additional Plots}\label{app:plots}
%\note[LS]{what was intended to go here? do we still need this appendix section?}

\begin{figure}[h]
    \centering
    \begin{minipage}{.48\textwidth}
    \includegraphics[width = \textwidth]{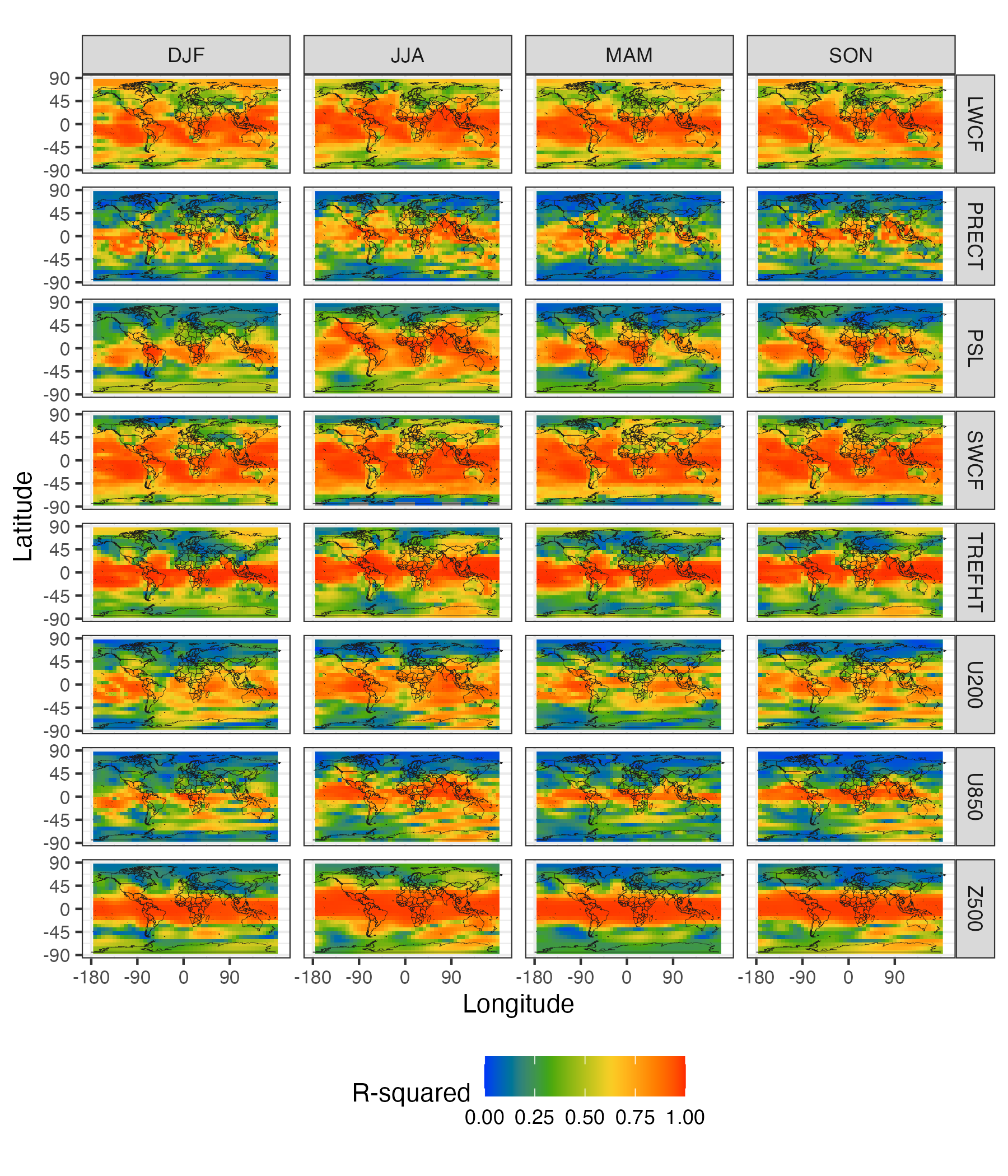}
\end{minipage}    \begin{minipage}{.48\textwidth}
        \includegraphics[width = \textwidth]{figures/SWCF_LWCF_PRECT_PSL_Z500_U200_U850_TREFHT_U_RELHUM_T_RESTOM_0.7_r2_lat_plev.png}

    \includegraphics[width = \textwidth]{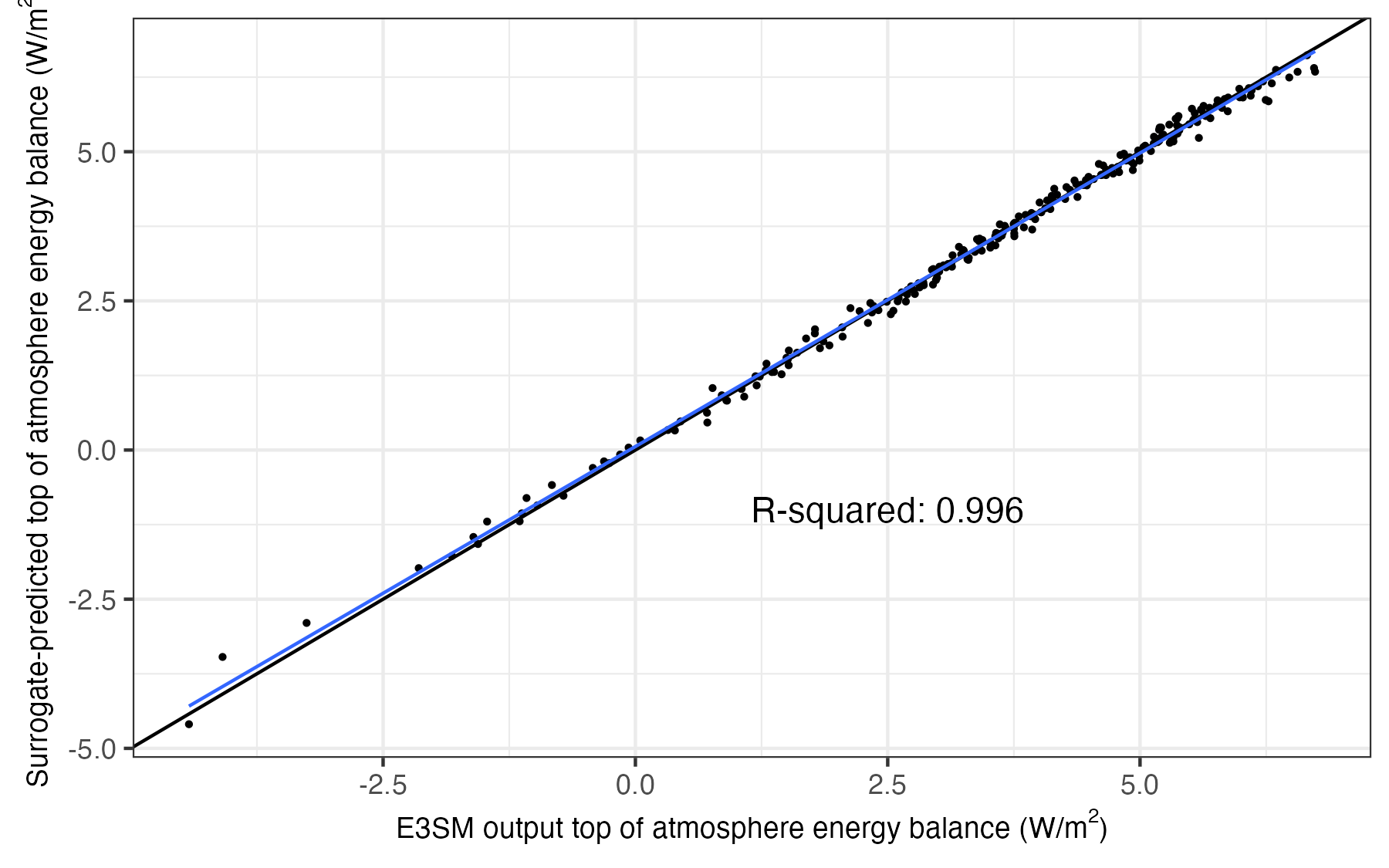}\end{minipage}
    \caption{Surrogate R-squared (proportion of explained variance) computed for each variables, season, and spatial location: (Left) for fields defined on longitude/latitude, (Top right) for fields defined on latitude/pressure, (Bottom right) a scatterplot of global top of atmosphere energy balance, with a black line at y=x and blue line representing the linear regression line between the simulation output and surrogate prediction using the 250 simulations. }
    \label{fig:rsquared_by_location_all}
\end{figure}

\begin{figure}[h]
    \centering
    \begin{minipage}{.48\textwidth}
    \includegraphics[width = \textwidth]{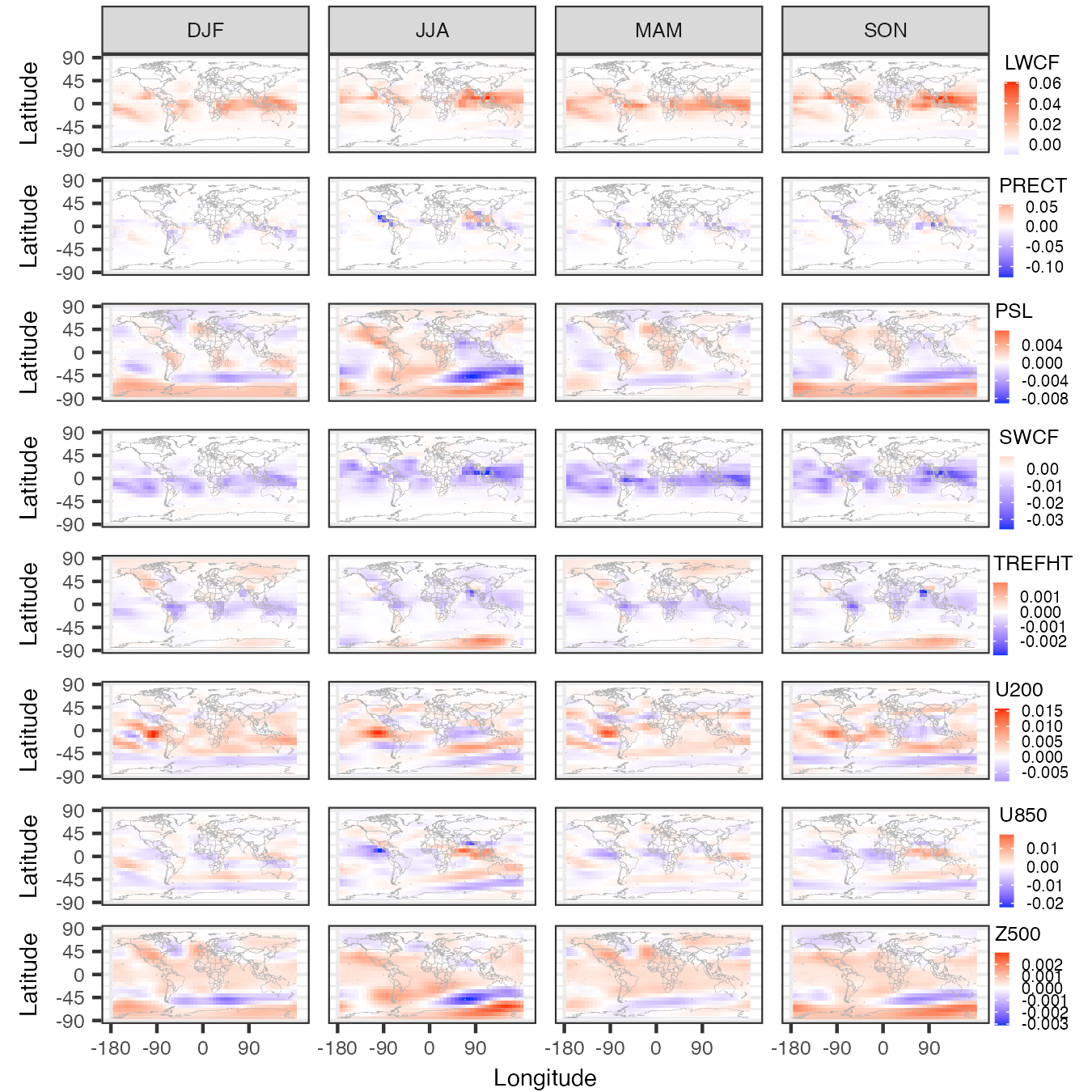}
\end{minipage}    \begin{minipage}{.48\textwidth}
        \includegraphics[width = \textwidth]{figures/SWCF_LWCF_PRECT_PSL_Z500_U200_U850_TREFHT_U_RELHUM_T_RESTOM_0.7_pc1_lat_plev.png}

    \includegraphics[width = \textwidth]{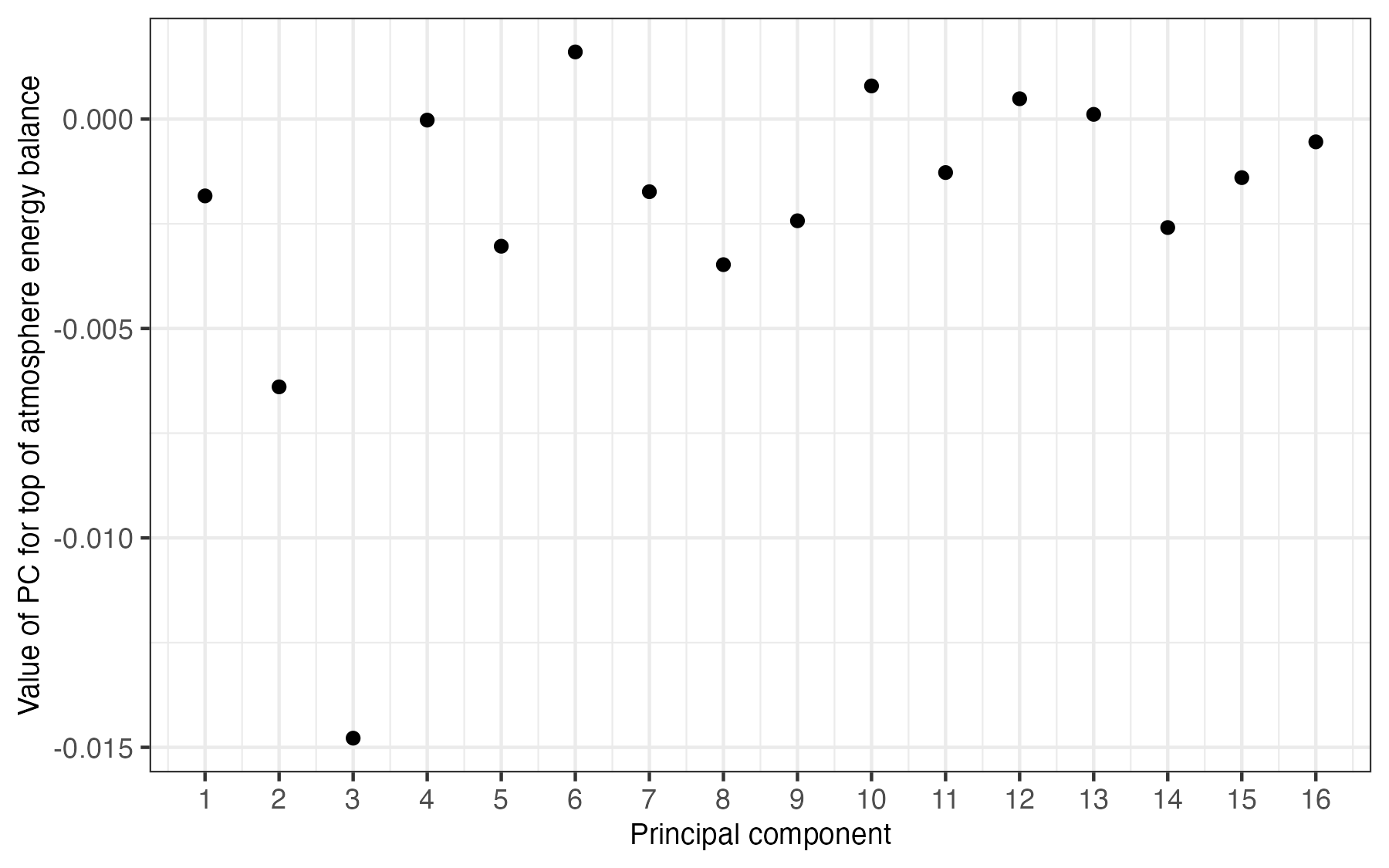}\end{minipage}
    \caption{Principal component (PC) vector values: (Left) for 3 fields defined on longitude/latitude and the first PC, (Top right) for fields defined on latitude/pressure for the first PC, (Bottom right) for global top of atmosphere energy balance for all PCs. }
    \label{fig:pc1_by_location_all}
\end{figure}

\begin{figure}[h]
    \centering
    \includegraphics[width = .48\textwidth]{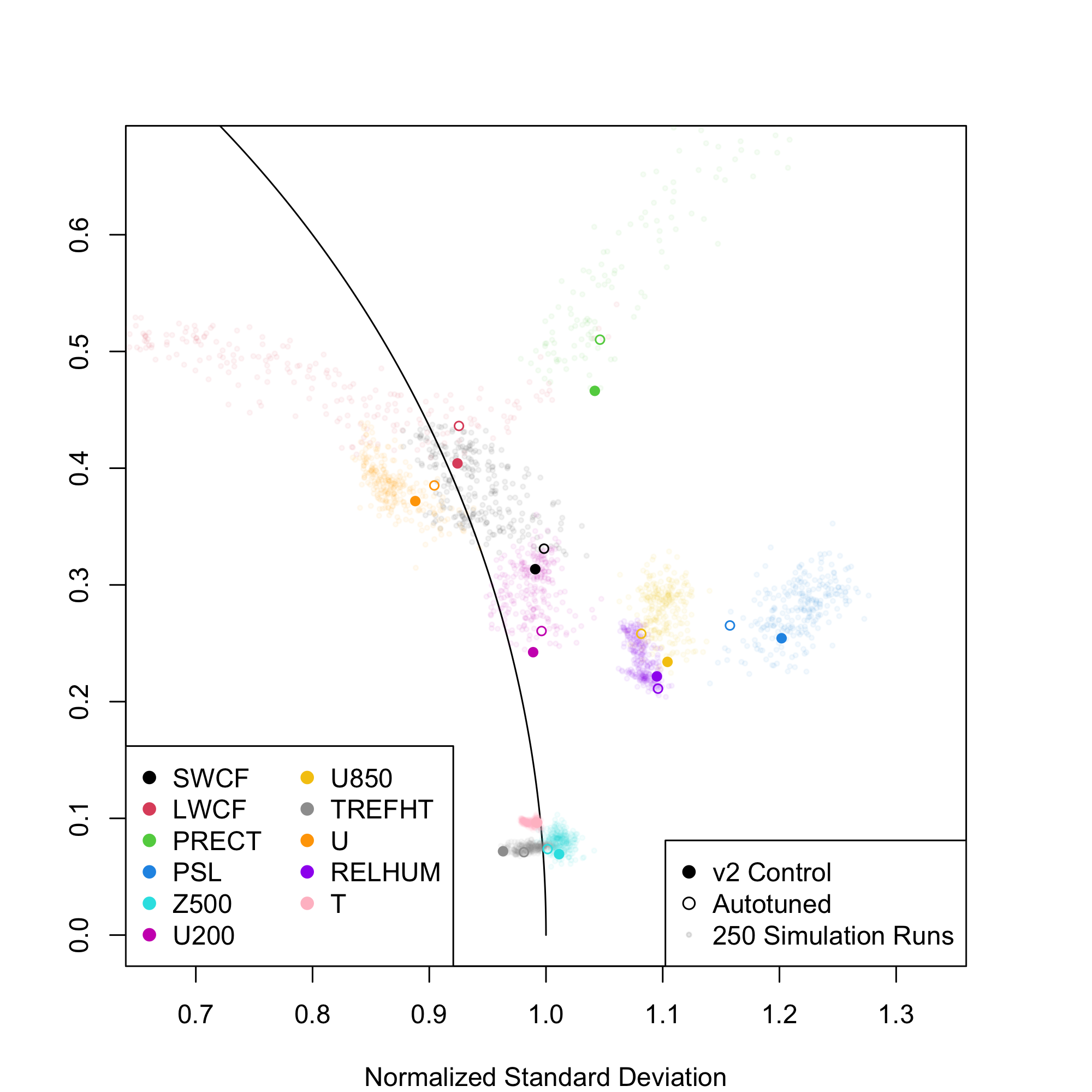}
    \includegraphics[width = .48\textwidth]{figures/taylor_MAM.png}

    \includegraphics[width = .48\textwidth]{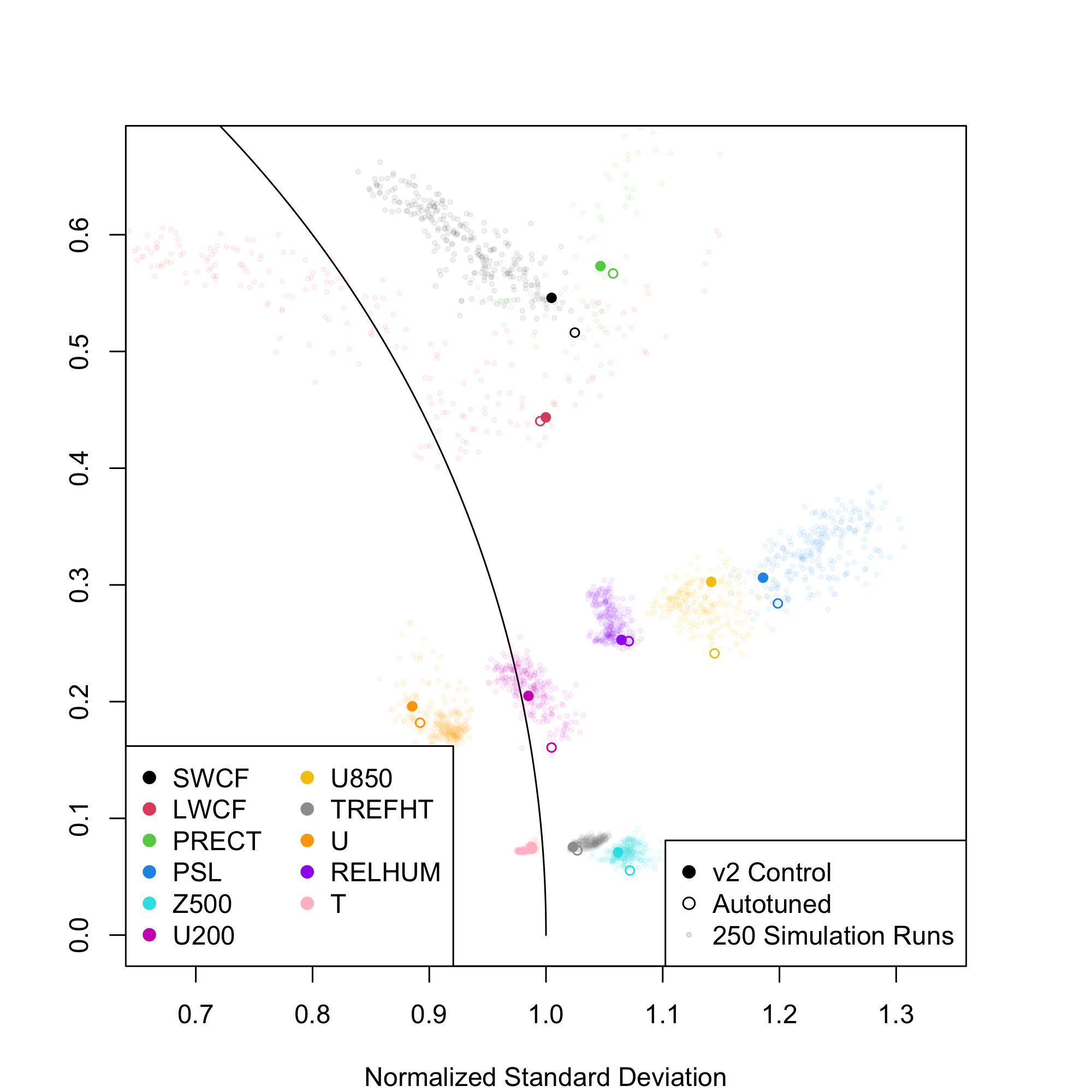}
    \includegraphics[width = .48\textwidth]{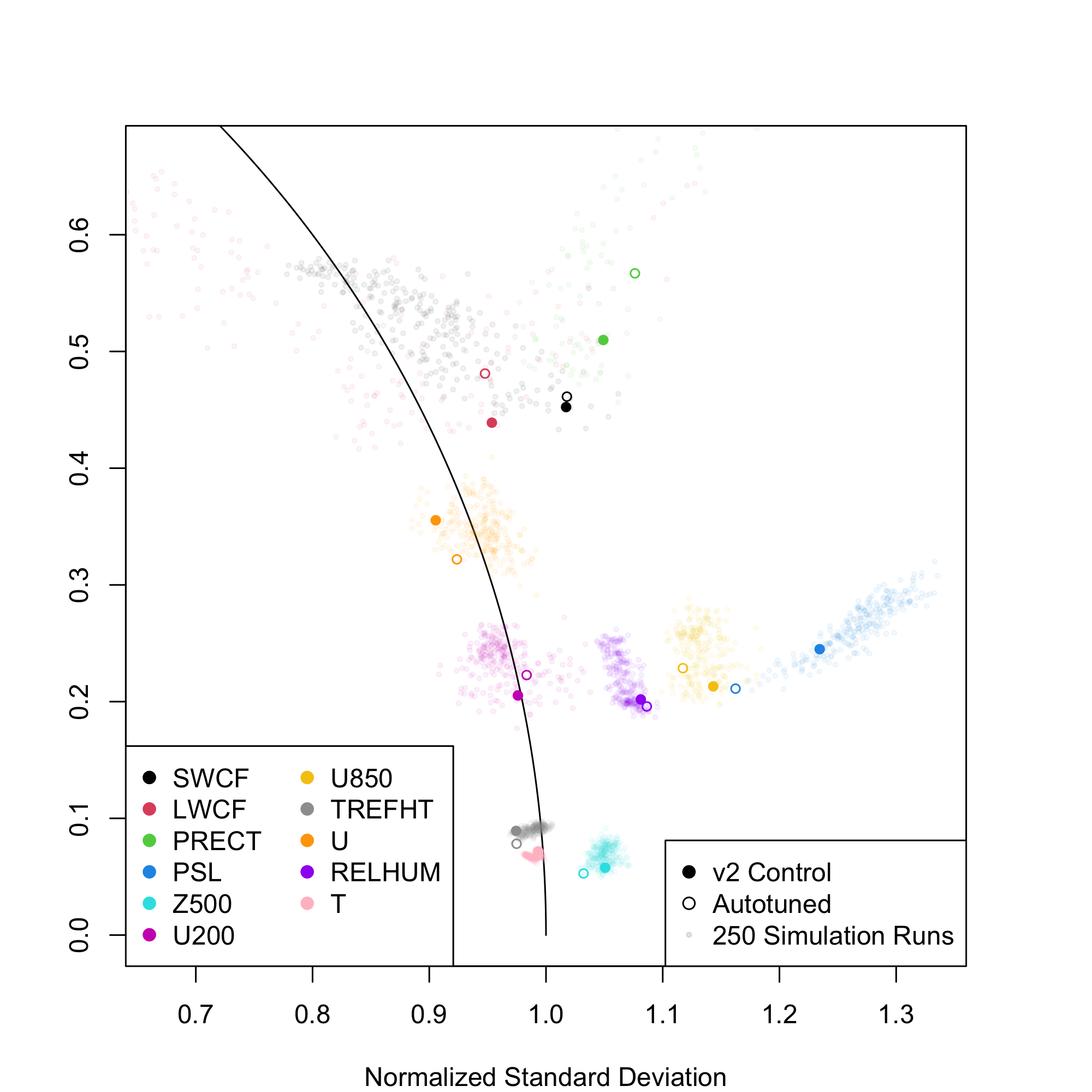}
    \caption{Taylor diagram of automatically-calibrated and control parameters for each season. (Top Left) DJF (Top Right) MAM (Bottom Left) JJA (Bottom Right) SON.}
    \label{fig:taylor_all}
\end{figure}

We present more complete plots from the surrogate results.
In figure \ref{fig:rsquared_by_location_all}, we provide the location-by-location R-squared, and show a plot of surrogate-predicted and model output RESTOM values. 
The second plot \ref{fig:pc1_by_location_all} shows the first principal component for all variables, including all principal components for RESTOM. 
In figure \ref{fig:taylor_all}, we provide Taylor diagrams for all four seasons, allowing for comparison between them.

\section{Open Research}

The E3SM code is available at \url{https://github.com/E3SM-Project/E3SM}. The exact E3SM code used to run the 250 training simulations can be retrieved by checking out the hash 37959275bf3384157264e45a8d9c7c43f2be1d56. The code pre-dates the official release of E3SMv2 but has the same climate, thus we refer to it as ``E3SMv2" in this publication. 

Our ensemble creation and surrogate code is available at \url{https://github.com/E3SM-Project/Autotuning-NGD}, which is currently closed to non-E3SM developers. We are working on making the code publicly available. Contact authors for code in the meantime. 
The code for the surrogate construction is based on the \texttt{scikit-learn} framework \cite{scikit-learn} and implemented in the \texttt{tesuract} package \cite{chowdhary_tesuract}. 
In addition, we use preprocessing code from the \texttt{clif} package at \url{https://github.com/sandialabs/clif}. 
Specifying surrogate and optimization options are done through YAML files. 
Our Python environment is based on the ``e3sm\_unified\_1.7.1\_chrysalis'' environment (\url{https://e3sm.org/resources/tools/other-tools/e3sm-unified-environment/}), and includes additional packages specified by the ``requirements.txt'' file on our GitHub repository. 
The code is also amenable to parallel architectures. 
All surrogate and optimization results were run on one node on Chrysalis.

The prescribed sea surface temperature and sea ice extent data is a 2005-2014 monthly climatology from \cite{sstdata}. 

%%%%%%%%%%%%%%%%%%%%%%%%%%%%%%%%%%%%%%%%%%%%%%%

\paragraph{Acknowledgments}
This paper describes objective technical results and analysis. Any subjective views or opinions that might be expressed in the paper do not necessarily represent the views of the U.S. Department of Energy or the United States Government.
Sandia National Laboratories is a multimission laboratory managed and operated by National Technology \& Engineering Solutions of Sandia, LLC, a wholly owned subsidiary of Honeywell International Inc., for the U.S. Department of Energy’s National Nuclear Security Administration under contract DE-NA0003525. SAND2023-XXX

We would like to acknowledge the helpful discussions with Andy Salinger (Sandia National Laboratories), Peter Caldwell (Lawrence Livermore National Laboratory) and Tommie Catanach (Sandia National Laboratories) and feedback from Claudia Tibaldi (Pacific Northwest National Laboratory), Bryce Harrop (Pacific Northwest National Laboratory), and Gavin Collins (Sandia National Laboratories)

The data was produced using a high-performance computing cluster provided by the BER Earth System Modeling program and operated by the Laboratory Computing Resource Center at Argonne National Laboratory.

\paragraph{Data Aailability}

Our post-processed ensemble and autocalibrated simulation data and our surrogate code are available at \url{https://github.com/E3SM-Project/AutoCalibration}.

\bibliographystyle{unsrtnat}
\bibliography{E3SMautotuning.bib}

\begin{thebibliography}{36}
\providecommand{\natexlab}[1]{#1}
\providecommand{\url}[1]{\texttt{#1}}
\expandafter\ifx\csname urlstyle\endcsname\relax
  \providecommand{\doi}[1]{doi: #1}\else
  \providecommand{\doi}{doi: \begingroup \urlstyle{rm}\Url}\fi

\bibitem[Hourdin et~al.(2017)Hourdin, Mauritsen, Gettelman, Golaz, Balaji,
  Duan, Folini, Ji, Klocke, Qian, Rauser, Rio, Tomassini, Watanabe, and
  Williamson]{hourdin_2017}
Frédéric Hourdin, Thorsten Mauritsen, Andrew Gettelman, Jean-Christophe
  Golaz, Venkatramani Balaji, Qingyun Duan, Doris Folini, Duoying Ji, Daniel
  Klocke, Yun Qian, Florian Rauser, Catherine Rio, Lorenzo Tomassini, Masahiro
  Watanabe, and Daniel Williamson.
\newblock The art and science of climate model tuning.
\newblock \emph{Bulletin of the American Meteorological Society}, 98\penalty0
  (3):\penalty0 589 -- 602, 2017.
\newblock \doi{https://doi.org/10.1175/BAMS-D-15-00135.1}.
\newblock URL
  \url{https://journals.ametsoc.org/view/journals/bams/98/3/bams-d-15-00135.1.xml}.

\bibitem[Jackson et~al.(2004)Jackson, Sen, and Stoffa]{jackson2004efficient}
Charles~S Jackson, Mrinal~K Sen, and Paul~L Stoffa.
\newblock An efficient stochastic {Bayesian} approach to optimal parameter and
  uncertainty estimation for climate model predictions.
\newblock \emph{Journal of Climate}, 17\penalty0 (14):\penalty0 2828--2841,
  2004.

\bibitem[Jackson et~al.(2008)Jackson, Sen, Huerta, Deng, and
  Bowman]{jackson2008error}
Charles~S Jackson, Mrinal~K Sen, Gabriel Huerta, Yi~Deng, and Kenneth~P Bowman.
\newblock Error reduction and convergence in climate prediction.
\newblock \emph{Journal of Climate}, 21\penalty0 (24):\penalty0 6698--6709,
  2008.

\bibitem[Kennedy and O'Hagan(2001)]{kennedy2001bayesian}
Marc~C Kennedy and Anthony O'Hagan.
\newblock Bayesian calibration of computer models.
\newblock \emph{Journal of the Royal Statistical Society: Series B (Statistical
  Methodology)}, 63\penalty0 (3):\penalty0 425--464, 2001.

\bibitem[Higdon et~al.(2013)Higdon, Gattiker, Lawrence, Jackson, Tobis,
  Pratola, Habib, Heitmann, and Price]{higdon2013computer}
Dave Higdon, Jim Gattiker, Earl Lawrence, Charles Jackson, Michael Tobis, Matt
  Pratola, Salman Habib, Katrin Heitmann, and Steve Price.
\newblock Computer model calibration using the ensemble {Kalman} filter.
\newblock \emph{Technometrics}, 55\penalty0 (4):\penalty0 488--500, 2013.

\bibitem[Dunbar et~al.(2021)Dunbar, Garbuno-Inigo, Schneider, and
  Stuart]{dunbar2021calibration}
Oliver~RA Dunbar, Alfredo Garbuno-Inigo, Tapio Schneider, and Andrew~M Stuart.
\newblock Calibration and uncertainty quantification of convective parameters
  in an idealized {GCM}.
\newblock \emph{Journal of Advances in Modeling Earth Systems}, 13\penalty0
  (9):\penalty0 e2020MS002454, 2021.

\bibitem[Berdahl et~al.(2021)Berdahl, Leguy, Lipscomb, and
  Urban]{berdahl2021statistical}
Mira Berdahl, Gunter Leguy, William~H Lipscomb, and Nathan~M Urban.
\newblock Statistical emulation of a perturbed basal melt ensemble of an ice
  sheet model to better quantify {Antarctic} sea level rise uncertainties.
\newblock \emph{The Cryosphere}, 15\penalty0 (6):\penalty0 2683--2699, 2021.

\bibitem[Beusch et~al.(2021)Beusch, Nicholls, Gudmundsson, Hauser, Meinshausen,
  and Seneviratne]{beusch2021emission}
Lea Beusch, Zebedee Nicholls, Lukas Gudmundsson, Mathias Hauser, Malte
  Meinshausen, and Sonia~I Seneviratne.
\newblock From emission scenarios to spatially resolved projections with a
  chain of computationally efficient emulators: {MAGICC} (v7. 5.1)—{MESMER}
  (v0. 8.1) coupling.
\newblock \emph{Geosci. Model Dev. Discuss}, 2021.

\bibitem[Cleary et~al.(2021)Cleary, Garbuno-Inigo, Lan, Schneider, and
  Stuart]{cleary2021calibrate}
Emmet Cleary, Alfredo Garbuno-Inigo, Shiwei Lan, Tapio Schneider, and Andrew~M
  Stuart.
\newblock Calibrate, emulate, sample.
\newblock \emph{Journal of Computational Physics}, 424:\penalty0 109716, 2021.

\bibitem[Fletcher et~al.(2022)Fletcher, McNally, Virgin, and
  King]{fletcher2022toward}
Christopher~G Fletcher, William McNally, John~G Virgin, and Fraser King.
\newblock Toward efficient calibration of higher-resolution {Earth} {System}
  {Models}.
\newblock \emph{Journal of Advances in Modeling Earth Systems}, 14\penalty0
  (7):\penalty0 e2021MS002836, 2022.

\bibitem[Ricciuto et~al.(2018)Ricciuto, Sargsyan, and
  Thornton]{ricciuto2018impact}
Daniel Ricciuto, Khachik Sargsyan, and Peter Thornton.
\newblock The impact of parametric uncertainties on biogeochemistry in the
  {E3SM} land model.
\newblock \emph{Journal of Advances in Modeling Earth Systems}, 10\penalty0
  (2):\penalty0 297--319, 2018.

\bibitem[Golaz et~al.(2022)Golaz, Van~Roekel, Zheng, Roberts, Wolfe, Lin,
  Bradley, Tang, Maltrud, Forsyth, Zhang, Zhou, Zhang, Zender, Wu, Wang,
  Turner, Singh, Richter, Qin, Petersen, Mametjanov, Ma, Larson, Krishna, Keen,
  Jeffery, Hunke, Hannah, Guba, Griffin, Feng, Engwirda, Di~Vittorio, Dang,
  Conlon, Chen, Brunke, Bisht, Benedict, Asay-Davis, Zhang, Zhang, Zeng, Xie,
  Wolfram, Vo, Veneziani, Tesfa, Sreepathi, Salinger, Reeves~Eyre, Prather,
  Mahajan, Li, Jones, Jacob, Huebler, Huang, Hillman, Harrop, Foucar, Fang,
  Comeau, Caldwell, Bartoletti, Balaguru, Taylor, McCoy, Leung, and
  Bader]{e3smv2overview}
Jean-Christophe Golaz, Luke~P. Van~Roekel, Xue Zheng, Andrew~F. Roberts,
  Jonathan~D. Wolfe, Wuyin Lin, Andrew~M. Bradley, Qi~Tang, Mathew~E. Maltrud,
  Ryan~M. Forsyth, Chengzhu Zhang, Tian Zhou, Kai Zhang, Charles~S. Zender,
  Mingxuan Wu, Hailong Wang, Adrian~K. Turner, Balwinder Singh, Jadwiga~H.
  Richter, Yi~Qin, Mark~R. Petersen, Azamat Mametjanov, Po-Lun Ma, Vincent~E.
  Larson, Jayesh Krishna, Noel~D. Keen, Nicole Jeffery, Elizabeth~C. Hunke,
  Walter~M. Hannah, Oksana Guba, Brian~M. Griffin, Yan Feng, Darren Engwirda,
  Alan~V. Di~Vittorio, Cheng Dang, LeAnn~M. Conlon, Chih-Chieh-Jack Chen,
  Michael~A. Brunke, Gautam Bisht, James~J. Benedict, Xylar~S. Asay-Davis,
  Yuying Zhang, Meng Zhang, Xubin Zeng, Shaocheng Xie, Phillip~J. Wolfram, Tom
  Vo, Milena Veneziani, Teklu~K. Tesfa, Sarat Sreepathi, Andrew~G. Salinger,
  J.~E.~Jack Reeves~Eyre, Michael~J. Prather, Salil Mahajan, Qing Li, Philip~W.
  Jones, Robert~L. Jacob, Gunther~W. Huebler, Xianglei Huang, Benjamin~R.
  Hillman, Bryce~E. Harrop, James~G. Foucar, Yilin Fang, Darin~S. Comeau,
  Peter~M. Caldwell, Tony Bartoletti, Karthik Balaguru, Mark~A. Taylor,
  Renata~B. McCoy, L.~Ruby Leung, and David~C. Bader.
\newblock The {DOE} {E3SM} model version 2: Overview of the physical model and
  initial model evaluation.
\newblock \emph{Journal of Advances in Modeling Earth Systems}, 14\penalty0
  (12), 2022.
\newblock \doi{https://doi.org/10.1029/2022MS003156}.
\newblock URL
  \url{https://agupubs.onlinelibrary.wiley.com/doi/abs/10.1029/2022MS003156}.

\bibitem[Durack and Taylor(2018)]{sstdata}
Paul~J. Durack and Karl~E. Taylor.
\newblock {PCMDI} {AMIP} {SST} and sea-ice boundary conditions version 1.1.4,
  2018.
\newblock URL \url{https://doi.org/10.22033/ESGF/input4MIPs.2204}.

\bibitem[McKay et~al.(1979)McKay, Beckman, and Conover]{McKay1979}
M.~D. McKay, R.~J. Beckman, and W.~J. Conover.
\newblock Comparison of three methods for selecting values of input variables
  in the analysis of output from a computer code.
\newblock \emph{Technometrics}, 21\penalty0 (2):\penalty0 239--245, 1979.
\newblock \doi{10.1080/00401706.1979.10489755}.

\bibitem[Golaz et~al.(2002)Golaz, Larson, and Cotton]{Golaz2002clubb}
Jean-Christophe Golaz, Vincent~E. Larson, and William~R. Cotton.
\newblock A {PDF}-based model for boundary layer clouds. {Part} {I}: {Method}
  and model description.
\newblock \emph{Journal of the Atmospheric Sciences}, 59\penalty0
  (24):\penalty0 3540 -- 3551, 2002.
\newblock
  \doi{https://doi.org/10.1175/1520-0469(2002)059<3540:APBMFB>2.0.CO;2}.
\newblock URL
  \url{https://journals.ametsoc.org/view/journals/atsc/59/24/1520-0469_2002_059_3540_apbmfb_2.0.co_2.xml}.

\bibitem[Larson and Golaz(2005)]{larson_clubb}
Vincent~E. Larson and Jean-Christophe Golaz.
\newblock Using probability density functions to derive consistent closure
  relationships among higher-order moments.
\newblock \emph{Monthly Weather Review}, 133\penalty0 (4):\penalty0 1023 --
  1042, 2005.
\newblock \doi{https://doi.org/10.1175/MWR2902.1}.
\newblock URL
  \url{https://journals.ametsoc.org/view/journals/mwre/133/4/mwr2902.1.xml}.

\bibitem[Zhang and McFarlane(1995)]{zm95}
G.J. Zhang and Norman~A. McFarlane.
\newblock Sensitivity of climate simulations to the parameterization of cumulus
  convection in the {Canadian} climate centre general circulation model.
\newblock \emph{Atmosphere-Ocean}, 33\penalty0 (3):\penalty0 407--446, 1995.
\newblock \doi{10.1080/07055900.1995.9649539}.
\newblock URL \url{https://doi.org/10.1080/07055900.1995.9649539}.

\bibitem[Xie et~al.(2019)Xie, Wang, Lin, Ma, Tang, Tang, Zheng, Golaz, Zhang,
  and Zhang]{xie2019}
Shaocheng Xie, Yi-Chi Wang, Wuyin Lin, Hsi-Yen Ma, Qi~Tang, Shuaiqi Tang, Xue
  Zheng, Jean-Christophe Golaz, Guang~J. Zhang, and Minghua Zhang.
\newblock Improved diurnal cycle of precipitation in {E3SM} with a revised
  convective triggering function.
\newblock \emph{Journal of Advances in Modeling Earth Systems}, 11\penalty0
  (7):\penalty0 2290--2310, 2019.
\newblock \doi{https://doi.org/10.1029/2019MS001702}.
\newblock URL
  \url{https://agupubs.onlinelibrary.wiley.com/doi/abs/10.1029/2019MS001702}.

\bibitem[Gettelman and Morrison(2015)]{MG2}
A.~Gettelman and H.~Morrison.
\newblock Advanced two-moment bulk microphysics for global models. {Part} {I}:
  {Off}-line tests and comparison with other schemes.
\newblock \emph{Journal of Climate}, 28\penalty0 (3):\penalty0 1268 -- 1287,
  2015.
\newblock \doi{https://doi.org/10.1175/JCLI-D-14-00102.1}.
\newblock URL
  \url{https://journals.ametsoc.org/view/journals/clim/28/3/jcli-d-14-00102.1.xml}.

\bibitem[Qian et~al.(2018)Qian, Wan, Yang, Golaz, Harrop, Hou, Larson, Leung,
  Lin, Lin, Ma, Ma, Rasch, Singh, Wang, Xie, and Zhang]{Qian2018}
Yun Qian, Hui Wan, Ben Yang, Jean-Christophe Golaz, Bryce Harrop, Zhangshuan
  Hou, Vincent~E. Larson, L.~Ruby Leung, Guangxing Lin, Wuyin Lin, Po-Lun Ma,
  Hsi-Yen Ma, Phil Rasch, Balwinder Singh, Hailong Wang, Shaocheng Xie, and Kai
  Zhang.
\newblock Parametric sensitivity and uncertainty quantification in the version
  1 of {E3SM} atmosphere model based on short perturbed parameter ensemble
  simulations.
\newblock \emph{Journal of Geophysical Research: Atmospheres}, 123\penalty0
  (23):\penalty0 13,046--13,073, 2018.
\newblock \doi{https://doi.org/10.1029/2018JD028927}.
\newblock URL
  \url{https://agupubs.onlinelibrary.wiley.com/doi/abs/10.1029/2018JD028927}.

\bibitem[Adams et~al.(2018)Adams, Bohnhoff, Dalbey, Ebeida, Eddy, Eldred,
  Geraci, Hooper, Hough, Hu, Jakeman, Khalil, Maupin, Monschke, Ridgway,
  Rushdi, Stephens, Swiler, Vigil, Wildey, , and Winokur]{dakota}
B.M. Adams, W.J. Bohnhoff, K.R. Dalbey, M.S. Ebeida, J.P. Eddy, M.S. Eldred,
  G.~Geraci, R.W. Hooper, P.D. Hough, K.T Hu, J.D. Jakeman, M.~Khalil, K.A.
  Maupin, J.A. Monschke, E.M. Ridgway, A.A. Rushdi, J.A. Stephens, L.P. Swiler,
  D.M. Vigil, T.M. Wildey, , and J.G. Winokur.
\newblock Dakota, a multilevel parallel object-oriented framework for design
  optimization, parameter estimation, uncertainty quantification, and
  sensitivity analysis: Version 6.8 user’s manual.
\newblock \emph{Sandia Technical Report SAND2014-4633}, May 2018.

\bibitem[Chowdhary et~al.(2022)Chowdhary, Hoang, Lee, Ray, Weirs, and
  Carnes]{Chowdhary2022}
Kenny Chowdhary, Chi Hoang, Kookjin Lee, Jaideep Ray, V.G. Weirs, and Brian
  Carnes.
\newblock Calibrating hypersonic turbulence flow models with the {HIFiRE}-1
  experiment using data-driven machine-learned models.
\newblock \emph{Computer Methods in Applied Mechanics and Engineering},
  401:\penalty0 115396, November 2022.
\newblock \doi{10.1016/j.cma.2022.115396}.
\newblock URL \url{https://doi.org/10.1016/j.cma.2022.115396}.

\bibitem[Xiu and Karniadakis(2002)]{Xiu2002}
Dongbin Xiu and George~Em Karniadakis.
\newblock The {Wiener}--{Askey} polynomial chaos for stochastic differential
  equations.
\newblock \emph{{SIAM} Journal on Scientific Computing}, 24\penalty0
  (2):\penalty0 619--644, January 2002.
\newblock \doi{10.1137/s1064827501387826}.
\newblock URL \url{https://doi.org/10.1137/s1064827501387826}.

\bibitem[Xiu and Hesthaven(2005)]{Xiu2005}
Dongbin Xiu and Jan~S. Hesthaven.
\newblock High-order collocation methods for differential equations with random
  inputs.
\newblock \emph{{SIAM} Journal on Scientific Computing}, 27\penalty0
  (3):\penalty0 1118--1139, January 2005.
\newblock \doi{10.1137/040615201}.
\newblock URL \url{https://doi.org/10.1137/040615201}.

\bibitem[Hesthaven et~al.(2007)Hesthaven, Gottlieb, and
  Gottlieb]{Hesthaven2007}
Jan~S. Hesthaven, Sigal Gottlieb, and David Gottlieb.
\newblock \emph{Spectral Methods for Time-Dependent Problems}.
\newblock Cambridge University Press, January 2007.
\newblock \doi{10.1017/cbo9780511618352}.
\newblock URL \url{https://doi.org/10.1017/cbo9780511618352}.

\bibitem[Bergstra and Bengio(2012)]{bergstra2012}
James Bergstra and Yoshua Bengio.
\newblock Random search for hyper-parameter optimization.
\newblock \emph{Journal of machine learning research}, 13\penalty0 (2), 2012.

\bibitem[Friedman et~al.(2010)Friedman, Hastie, and Tibshirani]{Friedman2010}
Jerome Friedman, Trevor Hastie, and Robert Tibshirani.
\newblock Regularization paths for generalized linear models via coordinate
  descent.
\newblock \emph{Journal of Statistical Software}, 33\penalty0 (1), 2010.
\newblock \doi{10.18637/jss.v033.i01}.
\newblock URL \url{https://doi.org/10.18637/jss.v033.i01}.

\bibitem[Chowdhary(2021)]{chowdhary_tesuract}
Kenny Chowdhary.
\newblock tesuract v.1.0, version 1.0, 10 2021.
\newblock URL \url{https://www.osti.gov/biblio/1861275}.

\bibitem[Pedregosa et~al.(2011)Pedregosa, Varoquaux, Gramfort, Michel, Thirion,
  Grisel, Blondel, Prettenhofer, Weiss, Dubourg, Vanderplas, Passos,
  Cournapeau, Brucher, Perrot, and Duchesnay]{scikit-learn}
F.~Pedregosa, G.~Varoquaux, A.~Gramfort, V.~Michel, B.~Thirion, O.~Grisel,
  M.~Blondel, P.~Prettenhofer, R.~Weiss, V.~Dubourg, J.~Vanderplas, A.~Passos,
  D.~Cournapeau, M.~Brucher, M.~Perrot, and E.~Duchesnay.
\newblock Scikit-learn: Machine learning in {P}ython.
\newblock \emph{Journal of Machine Learning Research}, 12:\penalty0 2825--2830,
  2011.

\bibitem[Byrd et~al.(1995)Byrd, Lu, Nocedal, and Zhu]{byrd1995limited}
Richard~H Byrd, Peihuang Lu, Jorge Nocedal, and Ciyou Zhu.
\newblock A limited memory algorithm for bound constrained optimization.
\newblock \emph{SIAM Journal on scientific computing}, 16\penalty0
  (5):\penalty0 1190--1208, 1995.

\bibitem[Gelman et~al.(1995)Gelman, Carlin, Stern, and
  Rubin]{gelman1995bayesian}
Andrew Gelman, John~B Carlin, Hal~S Stern, and Donald~B Rubin.
\newblock \emph{Bayesian data analysis}.
\newblock Chapman and Hall/CRC, 1995.

\bibitem[Ma et~al.(2022)Ma, Harrop, Larson, Neale, Gettelman, Morrison, Wang,
  Zhang, Klein, Zelinka, Zhang, Qian, Yoon, Jones, Huang, Tai, Singh,
  Bogenschutz, Zheng, Lin, Quaas, Chepfer, Brunke, Zeng, M\"ulmenst\"adt,
  Hagos, Zhang, Song, Liu, Pritchard, Wan, Wang, Tang, Caldwell, Fan, Berg,
  Fast, Taylor, Golaz, Xie, Rasch, and Leung]{ma2022}
P.-L. Ma, B.~E. Harrop, V.~E. Larson, R.~B. Neale, A.~Gettelman, H.~Morrison,
  H.~Wang, K.~Zhang, S.~A. Klein, M.~D. Zelinka, Y.~Zhang, Y.~Qian, J.-H. Yoon,
  C.~R. Jones, M.~Huang, S.-L. Tai, B.~Singh, P.~A. Bogenschutz, X.~Zheng,
  W.~Lin, J.~Quaas, H.~Chepfer, M.~A. Brunke, X.~Zeng, J.~M\"ulmenst\"adt,
  S.~Hagos, Z.~Zhang, H.~Song, X.~Liu, M.~S. Pritchard, H.~Wan, J.~Wang,
  Q.~Tang, P.~M. Caldwell, J.~Fan, L.~K. Berg, J.~D. Fast, M.~A. Taylor, J.-C.
  Golaz, S.~Xie, P.~J. Rasch, and L.~R. Leung.
\newblock Better calibration of cloud parameterizations and subgrid effects
  increases the fidelity of the e3sm atmosphere model version 1.
\newblock \emph{Geoscientific Model Development}, 15\penalty0 (7):\penalty0
  2881--2916, 2022.
\newblock \doi{10.5194/gmd-15-2881-2022}.
\newblock URL \url{https://gmd.copernicus.org/articles/15/2881/2022/}.

\bibitem[Taylor(2001)]{taylor_2001}
Karl~E. Taylor.
\newblock Summarizing multiple aspects of model performance in a single
  diagram.
\newblock \emph{Journal of Geophysical Research: Atmospheres}, 106\penalty0
  (D7):\penalty0 7183--7192, 2001.
\newblock \doi{https://doi.org/10.1029/2000JD900719}.
\newblock URL
  \url{https://agupubs.onlinelibrary.wiley.com/doi/abs/10.1029/2000JD900719}.

\bibitem[Lu et~al.(2015)Lu, Morzfeld, Tu, and Chorin]{lu2015limitations}
Fei Lu, Matthias Morzfeld, Xuemin Tu, and Alexandre~J Chorin.
\newblock Limitations of polynomial chaos expansions in the {Bayesian} solution
  of inverse problems.
\newblock \emph{Journal of Computational Physics}, 282:\penalty0 138--147,
  2015.

\bibitem[Yue et~al.(2014)Yue, Simpson, Lindgren, and Rue]{yue2014bayesian}
Yu~Ryan Yue, Daniel Simpson, Finn Lindgren, and H{\aa}vard Rue.
\newblock Bayesian adaptive smoothing splines using stochastic differential
  equations.
\newblock \emph{Bayesian Analysis}, 9\penalty0 (2):\penalty0 397 -- 424, 2014.
\newblock \doi{10.1214/13-BA866}.
\newblock URL \url{https://doi.org/10.1214/13-BA866}.

\bibitem[Chipman et~al.(2010)Chipman, George, and McCulloch]{chipman2010bart}
Hugh~A. Chipman, Edward~I. George, and Robert~E. McCulloch.
\newblock {BART: Bayesian additive regression trees}.
\newblock \emph{The Annals of Applied Statistics}, 4\penalty0 (1):\penalty0 266
  -- 298, 2010.
\newblock \doi{10.1214/09-AOAS285}.
\newblock URL \url{https://doi.org/10.1214/09-AOAS285}.

\end{thebibliography}

\end{document}